\documentclass[times,twocolumn]{aastex63}

\usepackage{comment}
\usepackage{color}
\usepackage{mhchem}
\usepackage{subfig}
\usepackage{moresize}
\usepackage{savesym}
\savesymbol{tablenum}
\usepackage{siunitx}
\restoresymbol{SIX}{tablenum}

\received{September 16, 2019}
\revised{October 15, 2020}
\accepted{October 15, 2020}
\submitjournal{AJ}

\shorttitle{GPI Characterization of HD~206893~B}
\shortauthors{Ward-Duong et al.}
\graphicspath{{./}{figures/}}

\begin{document}

\title{Gemini Planet Imager Spectroscopy of the Dusty Substellar Companion HD~206893~B}

\correspondingauthor{K. Ward-Duong}
\email{kwduong@stsci.edu}

\author[0000-0002-4479-8291]{K. Ward-Duong}
\affil{Five College Astronomy Department, Amherst College, Amherst, MA, 01002 }
\affil{School of Earth and Space Exploration, Arizona State University, P.O. Box 871404, Tempe, AZ 85287, USA}

\author{J. Patience}
\affiliation{School of Earth and Space Exploration, Arizona State University, P.O. Box 871404, Tempe, AZ 85287, USA}

\author[0000-0002-7821-0695]{K. Follette}
\affiliation{Physics \& Astronomy Department, Amherst College, 21 Merrill Science Dr., Amherst, MA 01002}

\author[0000-0002-4918-0247]{R. J. De Rosa}
\affiliation{Kavli Institute for Particle Astrophysics and Cosmology, Stanford University, Stanford, CA, USA 94305}
\affiliation{European Southern Observatory, Alonso de C\'{o}rdova 3107, Vitacura, Santiago, Chile}

\author[0000-0003-0029-0258]{J. Rameau}
\affiliation{Univ. Grenoble Alpes/CNRS, IPAG, F-38000 Grenoble, France}
\affiliation{Institut de Recherche sur les Exoplan{\`e}tes, D{\'e}partement de Physique, Universit{\'e} de Montr{\'e}al, Montr{\'e}al QC, H3C 3J7, Canada}

\author[0000-0002-5251-2943]{M. Marley}
\affiliation{NASA Ames Research Center,  Mountain View, CA, USA 94035}

\author{D. Saumon}
\affiliation{Los Alamos National Laboratory, P.O. Box 1663, Los Alamos, NM 87545, USA}

\author[0000-0001-6975-9056]{E. L. Nielsen}
\affiliation{Kavli Institute for Particle Astrophysics and Cosmology, Stanford University, Stanford, CA, USA 94305}

\author[0000-0002-9246-5467]{A. Rajan}
\affiliation{Space Telescope Science Institute, Baltimore, MD, USA 21218}

\author[0000-0002-7162-8036]{A. Z. Greenbaum}
\affiliation{Department of Astronomy, University of Michigan, Ann Arbor MI, USA 48109}

\author{J. Lee}
\affiliation{Department of Physics and Astronomy, University of Georgia, Athens, GA, USA 30602}

\author[0000-0003-0774-6502]{J. J. Wang}
\altaffiliation{51 Pegasi b Fellow}
\affiliation{Department of Astronomy, California Institute of Technology, Pasadena, CA 91125, USA}
\affiliation{Department of Astronomy, University of California, Berkeley, CA 94720, USA}

\author[0000-0002-1483-8811]{I. Czekala}
\altaffiliation{NASA Hubble Fellowship Program Sagan Fellow}
\affiliation{Department of Astronomy, University of California, Berkeley, CA 94720, USA}
\affiliation{Kavli Institute for Particle Astrophysics and Cosmology, Stanford University, Stanford, CA 94305, USA}

\author[0000-0002-5092-6464]{G. Duch\^ene}
\affiliation{Department of Astronomy, University of California, Berkeley, CA 94720, USA}
\affiliation{Univ. Grenoble Alpes/CNRS, IPAG, F-38000 Grenoble, France}

\author[0000-0003-1212-7538]{B. Macintosh}
\affiliation{Kavli Institute for Particle Astrophysics and Cosmology, Stanford University, Stanford, CA 94305, USA}

\author[0000-0001-5172-7902]{S. Mark Ammons}
\affiliation{Lawrence Livermore National Laboratory, Livermore, CA 94551, USA}

\author[0000-0002-5407-2806]{V. P. Bailey}
\affiliation{Jet Propulsion Laboratory, California Institute of Technology, Pasadena, CA 91109, USA}

\author[0000-0002-7129-3002]{T. Barman}
\affiliation{Lunar and Planetary Laboratory, University of Arizona, Tucson AZ 85721, USA}

\author{J. Bulger}
\affiliation{Institute for Astronomy, University of Hawaii, 2680 Woodlawn Drive, Honolulu, HI 96822, USA}
\affiliation{Subaru Telescope, NAOJ, 650 North A{'o}hoku Place, Hilo, HI 96720, USA}

\author{C. Chen}
\affiliation{Space Telescope Science Institute, Baltimore, MD 21218, USA}

\author[0000-0001-6305-7272]{J. Chilcote}
\affiliation{Kavli Institute for Particle Astrophysics and Cosmology, Stanford University, Stanford, CA 94305, USA}
\affiliation{Department of Physics, University of Notre Dame, 225 Nieuwland Science Hall, Notre Dame, IN, 46556, USA}

\author[0000-0003-0156-3019]{T. Cotten}
\affiliation{Department of Physics and Astronomy, University of Georgia, Athens, GA 30602, USA}

\author{R. Doyon}
\affiliation{Institut de Recherche sur les Exoplan{\`e}tes, D{\'e}partement de Physique, Universit{\'e} de Montr{\'e}al, Montr{\'e}al QC, H3C 3J7, Canada}

\author[0000-0002-0792-3719]{T. M. Esposito}
\affiliation{Department of Astronomy, University of California, Berkeley, CA 94720, USA}

\author[0000-0002-0176-8973]{M. P. Fitzgerald}
\affiliation{Department of Physics \& Astronomy, University of California, Los Angeles, CA 90095, USA}

\author[0000-0003-3978-9195]{B. L. Gerard}
\affiliation{Department of Astronomy, UC Santa Cruz, 1156 High St., Santa Cruz, CA 95064, USA}
\affiliation{National Research Council of Canada Herzberg, 5071 West Saanich Rd, Victoria, BC, V9E 2E7, Canada}

\author[0000-0002-4144-5116]{S. J. Goodsell}
\affiliation{Gemini Observatory, 670 N. A'ohoku Place, Hilo, HI 96720, USA}

\author{J. R. Graham}
\affiliation{Department of Astronomy, University of California, Berkeley, CA 94720, USA}

\author[0000-0003-3726-5494]{P. Hibon}
\affiliation{European Southern Observatory, Alonso de C\'{o}rdova 3107, Vitacura, Santiago, Chile}

\author[0000-0001-9994-2142]{J. Hom}
\affiliation{School of Earth and Space Exploration, Arizona State University, P.O. Box 871404, Tempe, AZ 85287, USA}

\author[0000-0003-1498-6088]{L.-W. Hung}
\affiliation{Natural Sounds and Night Skies Division, National Park Service, Fort Collins, CO 80525, USA}

\author{P. Ingraham}
\affiliation{Large Synoptic Survey Telescope, 950N Cherry Ave., Tucson, AZ 85719, USA}

\author[0000-0002-6221-5360]{P. Kalas}
\affiliation{Department of Astronomy, University of California, Berkeley, CA 94720, USA}
\affiliation{SETI Institute, Carl Sagan Center, 189 Bernardo Ave.,  Mountain View CA 94043, USA}

\author[0000-0002-9936-6285]{Q. Konopacky}
\affiliation{Center for Astrophysics and Space Science, University of California San Diego, La Jolla, CA 92093, USA}

\author{J. E. Larkin}
\affiliation{Department of Physics \& Astronomy, University of California, Los Angeles, CA 90095, USA}

\author{J. Maire}
\affiliation{Center for Astrophysics and Space Science, University of California San Diego, La Jolla, CA 92093, USA}

\author[0000-0001-7016-7277]{F. Marchis}
\affiliation{SETI Institute, Carl Sagan Center, 189 Bernardo Ave.,  Mountain View CA 94043, USA}

\author[0000-0002-4164-4182]{C. Marois}
\affiliation{National Research Council of Canada Herzberg, 5071 West Saanich Rd, Victoria, BC, V9E 2E7, Canada}
\affiliation{University of Victoria, 3800 Finnerty Rd, Victoria, BC, V8P 5C2, Canada}

\author[0000-0003-3050-8203]{S. Metchev}
\affiliation{Department of Physics and Astronomy, Centre for Planetary Science and Exploration, The University of Western Ontario, London, ON N6A 3K7, Canada}
\affiliation{Department of Physics and Astronomy, Stony Brook University, Stony Brook, NY 11794-3800, USA}

\author[0000-0001-6205-9233]{M. A. Millar-Blanchaer}
\affiliation{Jet Propulsion Laboratory, California Institute of Technology, Pasadena, CA 91109, USA}
\affiliation{NASA Hubble Fellow}

\author[0000-0001-7130-7681]{R. Oppenheimer}
\affiliation{Department of Astrophysics, American Museum of Natural History, New York, NY 10024, USA}

\author{D. Palmer}
\affiliation{Lawrence Livermore National Laboratory, Livermore, CA 94551, USA}

\author[0000-0002-3191-8151]{M. Perrin}
\affiliation{Space Telescope Science Institute, Baltimore, MD 21218, USA}

\author{L. Poyneer}
\affiliation{Lawrence Livermore National Laboratory, Livermore, CA 94551, USA}

\author{L. Pueyo}
\affiliation{Space Telescope Science Institute, Baltimore, MD 21218, USA}

\author[0000-0002-9667-2244]{F. T. Rantakyr\"o}
\affiliation{Gemini Observatory, Casilla 603, La Serena, Chile}

\author[0000-0003-1698-9696]{B. Ren}
\affiliation{Department of Physics and Astronomy, Johns Hopkins University, Baltimore, MD 21218, USA}

\author[0000-0003-2233-4821]{J.-B. Ruffio}
\affiliation{Kavli Institute for Particle Astrophysics and Cosmology, Stanford University, Stanford, CA 94305, USA}

\author[0000-0002-8711-7206]{D. Savransky}
\affiliation{Sibley School of Mechanical and Aerospace Engineering, Cornell University, Ithaca, NY 14853, USA}

\author[0000-0002-6294-5937]{A. C. Schneider}
\affiliation{US Naval Observatory, Flagstaff Station, P.O. Box 1149, Flagstaff, AZ 86002, USA}
\affiliation{Department of Physics and Astronomy, George Mason University, MS3F3, 4400 University Drive, Fairfax, VA 22030, USA}

\author[0000-0003-1251-4124]{A. Sivaramakrishnan}
\affiliation{Space Telescope Science Institute, Baltimore, MD 21218, USA}

\author[0000-0002-5815-7372]{I. Song}
\affiliation{Department of Physics and Astronomy, University of Georgia, Athens, GA 30602, USA}

\author[0000-0003-2753-2819]{R. Soummer}
\affiliation{Space Telescope Science Institute, Baltimore, MD 21218, USA}

\author{M. Tallis}
\affiliation{Kavli Institute for Particle Astrophysics and Cosmology, Stanford University, Stanford, CA 94305, USA}

\author{S. Thomas}
\affiliation{Large Synoptic Survey Telescope, 950N Cherry Ave., Tucson, AZ 85719, USA}

\author[0000-0001-5299-6899]{J. Kent Wallace}
\affiliation{Jet Propulsion Laboratory, California Institute of Technology, Pasadena, CA 91109, USA}

\author{S. Wiktorowicz}
\affiliation{The Aerospace Corporation, El Segundo, CA 90245, USA}

\author[0000-0002-9977-8255]{S. Wolff}
\affiliation{Leiden Observatory, Leiden University, P.O. Box 9513, 2300 RA Leiden, The Netherlands}




\begin{abstract}
We present new near-infrared Gemini Planet Imager (GPI) spectroscopy of HD~206893~B, a substellar companion orbiting within the debris disk of its F5V star. The \textit{J}, \textit{H}, \textit{K1}, and \textit{K2} spectra from GPI demonstrate the extraordinarily red colors of the object, confirming it as the reddest substellar object observed to date. The significant flux increase throughout the infrared presents a challenging atmosphere to model with existing grids. Best-fit values vary from 1200~K to 1800~K for effective temperature and from 3.0 to 5.0 for log($g$), depending on which individual wavelength band is fit and which model suite is applied. The extreme redness of the companion can be partially reconciled by invoking a high-altitude layer of sub-micron dust particles, similar to dereddening approaches applied to the peculiar red field L-dwarf population. However, reconciling the HD~206893~B spectra with even those of the reddest low-gravity L-dwarf spectra still requires the contribution of additional atmospheric dust, potentially due to the debris disk environment in which the companion resides. Orbit fitting from four years of astrometric monitoring is consistent with a $\sim$30-year period, orbital inclination of 147$^{\circ}$, and semimajor axis of 10~au, well within the estimated disk inner radius of $\sim$50~au. As one of very few substellar companions imaged interior to a circumstellar disk, the properties of this system offer important dynamical constraints on companion-disk interaction and provide a benchmark for substellar and planetary atmospheric study.
\end{abstract}




\section{Introduction} 
\label{sec:intro}

Low-mass directly-imaged companions, ranging from brown dwarfs to extrasolar giant planets, present an ideal laboratory to test formation mechanisms, measure atmospheric properties, and constrain the frequency of companions at large orbital separations. The population of directly-imaged giant planets remains small \cite[see review and recent survey results:][]{bowler16, galicher16, nielsen19}, and merits careful comparison with higher-mass brown dwarf companions observed over a wide range of system separations. Together with both free-floating planetary mass objects \citep[e.g.,][]{liu13, gagne15, kellogg16, schneider16} and field brown dwarfs \citep[e.g.,][]{burgasser06, kirkpatrick11}, substellar companions provide more readily characterizable analogs for study and comparison with planetary companions. 

Within the brown dwarf population itself, a diversity of spectral features, intrinsic color, and luminosity has been observed. Spectroscopic analysis has shown that objects with the same optical spectral classification may have divergent near-infrared (NIR) colors \citep{leggett03}, with younger objects occupying a significantly redder and fainter region of color-magnitude space and systematically cooler effective temperatures \citep[e.g.,][]{filippazzo15}. NIR color-magnitude analyses of field brown dwarfs, planetary-mass companions, and isolated low-mass, low-gravity objects have demonstrated distinct sequences between the three populations, corresponding to diversity in physical properties across spectral type, gravity, and age. \citet{liu16} identified similar photometric properties for late-M to late-L isolated (``free-floating'') substellar objects and young bound companions, but with indications that the young field population is brighter and/or redder (even for systems within the same moving groups). With the relative paucity of substellar companions, each discovery provides a unique opportunity for  comparison across various classes of substellar objects, contributing useful context that improves our understanding of brown dwarf/planet physical processes, formation pathways, and composition.

Bound substellar companion systems that also host a circumstellar disk provide opportunities to constrain and characterize the orbital properties of low-mass companions and disk-companion dynamical interaction. Studies with high-contrast imaging techniques have now imaged systems with companions in the region between the warm inner disk \citep[e.g., HR~3549;][]{mawet15} and the inner edge of a substantial outer debris disk \citep[e.g., HR~2562;][]{konopacky16}. While directly imaged companions can explain the observed disk morphology for systems like HR~2562, in other cases the presence of additional planets is needed to explain complex disk geometries \citep[e.g., HD~95086;][]{rameau16}. Even fewer systems have proven amenable to resolved imaging of both the companion and the disk simultaneously. With marginally-resolved \SI{70}{\um} \textit{Herschel} PACS imaging, faint disk structure from broad-band SPHERE imaging, and a close substellar companion at 11~au \citep{milli17}, the HD~206893 system is among only six systems hosting resolved disks and companions, in addition to $\beta$ Pictoris \citep{lagrange10}, Fomalhaut \citep{kalas08}, LkCa~15 \citep{thalmann10, krausireland12}, HD~106906 \citep{bailey14}, and PDS~70 \citep{keppler18}.

\subsection{HD~206893 System Properties}

Using the Spectro-Polarimetric High-contrast Exoplanet REsearch \citep[SPHERE;][]{beuzit08} instrument on VLT, the SPHERE High Angular Resolution Debris Disc Survey (SHARDDS) discovered a low-mass companion orbiting the F5V star HD~206893 \citep{milli17}. The star was initially selected as a target for the SHARDDS resolved debris disk study with the SPHERE/IRDIS instrument \citep{dohlen08} owing to the high fractional IR luminosity of its disk ($L_{dust}/L_{*} = 2.3 \times 10^{-4}$; \citealp{moor06}), and characterization of the spectral energy distribution (SED) from \citet{chen14}. After an initial $H$-band detection of the companion in 2015, subsequent follow-up in 2016 with VLT/NaCo $L^{'}$ imaging using an annular groove phase mask coronagraph \citep{mawet05} recovered the source at $L^{'}$. With a bright $L^{'} = 13.43^{+0.17}_{-0.15}$ magnitude relative to the initial measurement of $H = 16.79 \pm 0.06$, indicating an extremely red color, the object's position in the color-magnitude diagram is analogous to the L5--L9 field dwarf population. With two imaged epochs, \citet{milli17} confirmed the comoving nature of the object using an additional non-detection at the expected position of a background object in archival \textit{Hubble Space Telescope}/NICMOS data. Without known association to a moving group, and with literature ages ranging from 0.2--2.1~Gyr \citep{zuckermansong04, davidhillenbrand15}, companion mass estimates range from 24--73$M_\textnormal{Jup}$ using the COND models \citep{baraffe03}. With SPHERE IFS and IRDIS observations, \citet{delorme17} conducted a follow-up study of the system, including revised stellar properties of the primary (estimating solar metallicity and an age from 50--700~Myr), and analysis of $R\sim30$ $YJ$ spectra (\num{0.95} to \SI{1.64}{\um}) and \num{2.11} and \SI{2.25}{\um} $K$-band photometry of the companion. These data and analyses substantiated the unusual properties of the companion, establishing it as the reddest of any currently-known substellar objects, and provided an estimation of its spectral properties as those of a dusty, intermediate gravity L-dwarf. 

In addition to the brown dwarf detection, the debris disk of the system was tentatively detected in the widefield IRDIS data, and also marginally resolved in archival \textit{Herschel} \SI{70}{\um} data. From the \emph{Herschel} imaging, \citet{milli17} applied a joint SED and imaging fit to estimate the parameters of the debris disk, reporting an inner edge of 50~au, position angle of $60\pm10^{\circ}$, and inclination of $40\pm10^{\circ}$ from face-on. 

The orbital properties of HD~206893~B were revisited by \citet{grandjean19}, who performed a joint fit to radial velocity (RV) monitoring, \textit{Hipparcos} and \textit{Gaia} astrometry, and SPHERE and NaCo imaging to infer an orbital period for the companion of 21--33~yr. They derived its dynamical mass to be a potentially planetary mass of 10$^{+5}_{-4}M_\textnormal{Jup}$. However, \citet{grandjean19} note that the fit appears dominated by the \textit{Hipparcos-Gaia} proper motion constraints, as the estimated mass from the joint fit is inconsistent at the 2$\sigma$ level with the mass derived from the combination of only direct imaging astrometry and RV variation. The authors thus posit that the `B' companion cannot be responsible for the observed 1.6~year RV drift, and that this variation may correspond instead to an additional $\sim$15$M_\textnormal{Jup}$ interior companion with a short 1.6--4~year period.

In this study, we present Gemini Planet Imager \citep[GPI;][]{macintosh08} observations of the HD~206893 system at $J$, $H$, $K1$, and $K2$ bands with the goal of characterizing the imaged companion and providing the most comprehensive and highest-resolution spectral coverage to date. In Sections~\ref{sec:obs} and \ref{sec:reduce}, we detail the GPI observations, data reduction, post-processing, and analysis techniques. Assessment of the host star properties are described in Section~\ref{sec:stellar}. Results are described in Section~\ref{sec:results}, including companion photometry, location in the color-magnitude diagram, atmospheric characterization, astrometric analysis, and limits on additional companions. We examine potential explanations for the reddening of the system in Section~\ref{sec:discuss}, and conclude by comparing HD~206893~B with the known population of directly-imaged companions and disk systems in Section~\ref{sec:conclusion}.


\section{Observations} 
\label{sec:obs}

The Gemini Planet Imager Exoplanet Survey \citep[GPIES;][]{macintosh14} is a dedicated 900-hour direct imaging survey to discover and characterize giant planets within a sample of over 600 nearby young stars using the GPI integral field spectrograph (IFS) at Gemini South observatory. As part of the GPIES campaign, IFS coronagraphic observations of HD~206893 were first obtained in $H$-band spectroscopic mode on UT~2016-09-22 (program ID GS-2015B-Q-01). The target was included in the campaign sample owing to its proximity, youth, and strong infrared excess, and the companion was noted in the preliminary data reductions. A total of $38\times60$-second IFS datacubes were taken under atmospheric conditions with $\sim$ $0\farcs68$ average seeing, close to the median DIMM seeing for Gemini South ($\sim$ $0\farcs65$). Observations were taken near target transit at a nearly constant airmass, achieving a total of 32.7$^{\circ}$ of field rotation and enabling Angular Differential Imaging \citep[ADI;][]{marois06}. To obtain wavelength calibration and account for instrumental flexure, argon arc lamp frames were taken immediately preceding the science observations. Additional $H$-band observations were taken in polarization mode following the spectroscopic imaging sequence, but are not included in this analysis.

Follow-up spectroscopy was obtained on UT~2016-10-21 in $K1$ ($74\times60$-second frames; 59.7$^{\circ}$ rotation) and on UT~2016-11-17/18 in $K2$ (for a two-night total of $148\times60$-second frames; 29.5$^{\circ}$ rotation). Seeing conditions were worse than median in all three follow-up datasets, as shown in Table~\ref{tab:observations}. The object was re-observed in $K2$ on UT~2017-11-09 with the same observing methodology, achieving additional field rotation under improved atmospheric conditions with lower residual wavefront error. The object was observed once more to obtain a $J$ band spectrum on UT~2018-09-24. A summary of the instrument modes, observations, and seeing data is provided in Table~\ref{tab:observations}.


\begin{table}[]
\scriptsize
\centering
\begin{tabular}{ccccccc}
\hline
\hline
Night  & Band & $\lambda$/$\Delta\lambda$ & Int. Time & Field Rot. & Airmass & Seeing \\
(UT) & & & (min) & ($^{\circ}$) & Range & (arcsec)\\
\hline
2016-09-22       & $H$  & 44-49  & 37.7  & 32.7   & 1.05-1.06 & 0.68      \\
2016-10-21       & $K1$ & 62-70  & 73.5  & 59.7   & 1.05-1.09 & 1.17      \\
2016-11-17$^{*}$ & $K2$ & 75-83  & 75.5  & 14.9   & 1.12-1.54 & 1.46      \\
2016-11-18$^{*}$ & $K2$ & 75-83  & 71.5  & 14.6   & 1.12-1.40 & 1.44  \\    
2017-11-09       & $K2$ & 75-83  & 88    & 24.1   & 1.08-1.32 & $^{\dagger}$  \\          
2018-09-24       & $J$  & 35-39  & 89    & 68.1   & 1.05-1.11 & $^{\dagger}$  \\      
\hline
\end{tabular}
\caption{Observation Summary. Datasets denoted with $^{*}$ are shown for reference but not used in the modeling analyses presented in this work due to lower SNR. The $^{\dagger}$ denotes that facility seeing data were unavailable during the observations. \label{tab:observations}}
\end{table}



\section{Data Reduction and Analysis} 
\label{sec:reduce}

\subsection{GPI Data Reduction}

Calibrated spectral datacubes ($x, y, \lambda$) for datasets in each of the four bands were obtained from the raw IFS data using the GPI Data Reduction Pipeline v.1.4.0 \citep[DRP;][]{perrin16}. Within the GPI DRP, raw IFS frames are dark subtracted then interpolated over bad pixels. The effects of instrumental flexure on the positioning of individual microspectra in the IFS frames are accounted for by aligning the contemporaneous arc lamp frames against a library of deep reference argon arcs \citep{wolff14}, producing a wavelength calibration used to assemble spectral datacubes from the extracted microspectra \citep{maire14}. The resulting datacubes are then flatfielded, interpolated to a common wavelength axis, and corrected for geometric distortion \citep{konopacky14}, resulting in spectral datacubes with 37 channels each. The astrometric solution for GPI has been shown to remain constant within uncertainties \citep{konopacky14, derosa15}, and we adopt plate scale and position angle values of $14.166 \pm 0.007$ mas/pixel and $-0.10 \pm 0.13 ^{\circ}$E of N.

To align and register the images, the stellar position behind the GPI coronagraphic occulting mask was estimated using the four satellite spots present in each slice of the datacube \citep{wang14}. These spots are attenuated replica images of the stellar point spread function (PSF) introduced by the placement of a two-dimensional amplitude grating at the pupil plane \citep{marois06b, sivaramakrishnan06}, and enable estimation of the object flux to stellar flux ratio \citep{wang14}. The adopted spot-to-star flux ratios for each filter and associated apodizer were $1.798 \times 10^{-4}$ for $J$, $2.035 \times 10^{-4}$ for $H$, $2.695 \times 10^{-4}$ for $K1$, and $1.905 \times 10^{-4}$ for $K2$ \citep{maire14}.

Outside of the GPI~DRP, the fully-calibrated datacubes were further post-processed to subtract the contribution of the PSF and speckle noise. This was accomplished using three post-processing techniques: Locally-Optimized Combination of Images \citep[LOCI;][]{lafreniere07} for the $J$, $H$, and $K2$-band data, classical ADI \citep[cADI;][]{marois06} for the $K1$ data, and pyKLIP for all four bands \citep{wang15}. In this work, the pyKLIP reductions from the auto-reduced GPI pipeline \citep{wang18} are used to measure the achieved contrast and sensitivity to companions (see Section~\ref{sec:completeness} and the reduced pyKLIP images in Appendix~\ref{sec:appendix_pyklip_covar}), and the LOCI/cADI reductions are used for the companion spectral extraction and astrometry.

An apodized high pass filter following a Hanning profile (cutoff of $4$ equivalent pixels) was applied to each frame to suppress slowly varying spatial frequencies within the data that are not well captured by PSF subtraction algorithms. The small amount of field rotation at $H$ and increased speckle noise at $J$ and $K2$ prevent a clean extraction of the companion with cADI, so LOCI was used to further process these bands. In order to reproduce images optimized to match the spatial distribution of speckles, the best-matched PSFs for each band were then estimated from a library of reference images with the following routine parameters: a PSF subtraction annulus of $d_{r} = 5$ pixels, optimized region for PSF comparison of $300\times$ the full-width at half-maximum (FWHM), geometry factor of $g=1$, and separation criteria between regions of $N_{\delta} = 0.75$. For $K2$, the decreasing SNR of satellite spots near the red edge of the band, as well as the increased brightness of the companion, impacted the fidelity of the spectral extraction, requiring a modified high-pass filter with a cutoff of $10$ pixels. (Extractions of the $K2$ spectrum using various algorithms and filter sizes are presented in Appendix~\ref{sec:appendix_pyklip_covar}.) The $K1$ data benefit from a large field rotation, and simple cADI was used to perform speckle subtraction, subtracting the stellar halo contribution by taking the median of all science frames and subtracting the median from each individual frame.

\subsection{Astrometry, Spectroscopy, and Covariance}
\label{sec:reduce_spec}
The astrometry of the companion was measured using the collapsed broadband PSF-subtracted datacubes. To extract spectroscopy of the companion, the process of injecting a negative fake planet was applied \citep[cf.][]{marois10}. A negative template PSF with the estimated position and flux of the companion was inserted into the raw dataset, and then the post-processing pipelines applied above were repeated iteratively to minimize the residuals within a $2\times2$ FWHM region centered at the companion position, adjusting the x- and y-positions and flux of the companion in each iteration. This allowed for correction of any differences in algorithm throughput between the LOCI and cADI reductions. To correct for flux offsets between the $K1$ and  $K2$ datasets in the overlap region of the two bands (from \SI{2.10}{\um} -- \SI{2.20}{\um}), the $K1$ spectrum flux scaling was shifted to minimize the $\chi^{2}$ value between the two spectra, resulting in a 4\% downward shift. 

As the raw IFS data include over 36,000 microspectra, the process of spectral extraction and interpolation (from the 16 pixel microspectra to 37 wavelength elements) introduces correlations between the resulting 37 spectral channels in the final datacubes. Applying the methodology of \citet{grecobrandt16}, we calculate the covariance between channels to better estimate spectral uncertainties (see Appendix~\ref{sec:appendix_pyklip_covar}), an approach that has been applied to other recent analyses of GPI companion spectra \citep[cf.,][]{derosa16,johnson-groh17, rajan17}. This effect can then be incorporated into the error budget of the individual spectrophotometric points for model fitting, as described in Section~\ref{sec:modeling}.


\section{Host Star Properties} 
\label{sec:stellar}
\begin{figure}[h]
    \centering
    \includegraphics[width=0.48\textwidth]{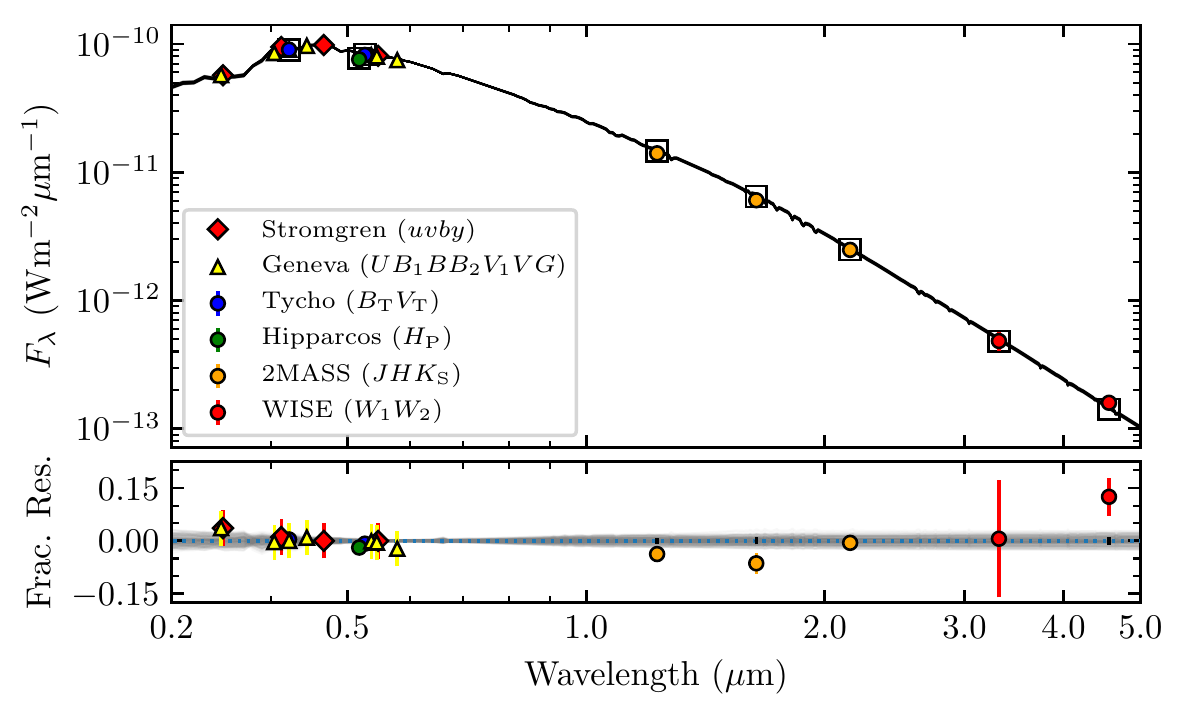}
    \caption{Stellar SED and best fit BT NextGen model spectrum for the primary star HD~206893~A, with residuals shown in the bottom panel. Best fit parameters correspond to intermediate age (601~Myr, albeit with large uncertainties of $^{+420}_{-380}$~Myr), and low extinction ($A_{V}=0.07^{+0.05}_{-0.04}$).}
    \label{fig:stellarsed}
\end{figure}

\begin{figure}[h]
    \centering
    \includegraphics[width=0.45\textwidth]{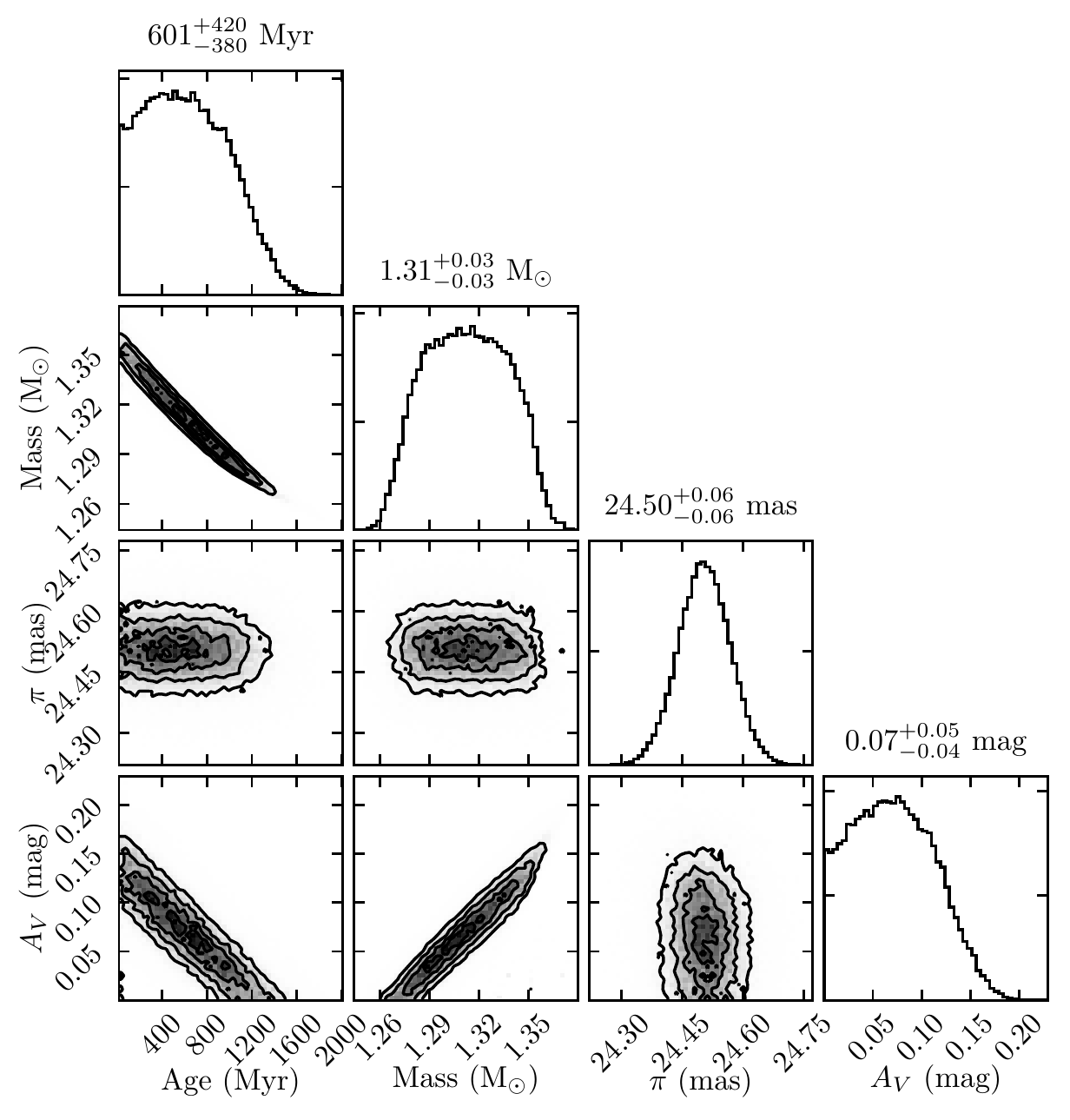}
    \caption{Posterior distribution of stellar properties of the primary star HD~206893~A. The asymmetric distribution on the stellar age suggests a broad range of $\sim$200~Myr to $\sim$1~Gyr at 1$\sigma$ confidence.}
    \label{fig:stellar_posteriors}
\end{figure}

We derive stellar properties from comparison of the stellar SED to BT-NextGen stellar model atmospheres \citep{allard12} combined with the MIST stellar isochrones \citep{dotter16, choi16}. For each stellar model grid at a given mass and age, stellar radius, temperature, and surface gravity are estimated and synthetic photometry is derived. Figure~\ref{fig:stellarsed} shows the combined Stromgren, Geneva, Tycho, \textit{Hipparcos}, 2MASS and \textit{WISE} photometry for HD~206893~A with the best-fit stellar atmospheric model overplotted. A Markov Chain Monte Carlo (MCMC) approach was applied to determine the best-fitting model, with stellar age, mass, extinction, and distance as free parameters \citep[following][]{nielsen17, derosa16}. Uniform priors in age, mass, and $A_{v}$ were applied, as well as a Gaussian distance prior centered at the observed \emph{Gaia} parallax. The metallicity is assumed to be solar, concordant with the findings of \citet{delorme17}. Figure~\ref{fig:stellar_posteriors} shows the posterior distribution of stellar properties from the MCMC analysis. Based upon the age posterior from these SED fits, we find that the stellar age is 601$^{+420}_{-380}$~Myr with 1$\sigma$ confidence. The best-fit stellar parameters are provided in Table~\ref{tab:starcomp_properties}. The large uncertainty on the stellar age from this approach is commensurate with the difficulty in determining precise age estimates for F5V stars, as noted in the detailed analysis of HD~206893~A by \citet{delorme17}. 

HD~206893~A is not a known member of a stellar association, which would allow for a more precise age determination. We cross-checked the possibility of young moving group membership using the Bayesian Analysis for Nearby Young AssociatioNs (BANYAN) $\Sigma$ tool \citep{gagne18}. Given \textit{Gaia}~DR2 values for stellar parallax, proper motion, and radial velocity\footnote{The RV drift of $-0.1\pm0.5$ km/s identified in \citet{grandjean19} is within the \textit{Gaia} DR2 uncertainty ($-12.45\pm0.59$ km/s) and likely does not affect this kinematic estimate.} \citep[Table~\ref{tab:starcomp_properties};][]{gaiadr2}, BANYAN~$\Sigma$ provides a $61.1\%$ probability of membership in the Argus moving group, and a $38.9\%$ probability of field membership.  

\begin{figure*}[h]
    \centering
    \includegraphics[width=1.0\textwidth]{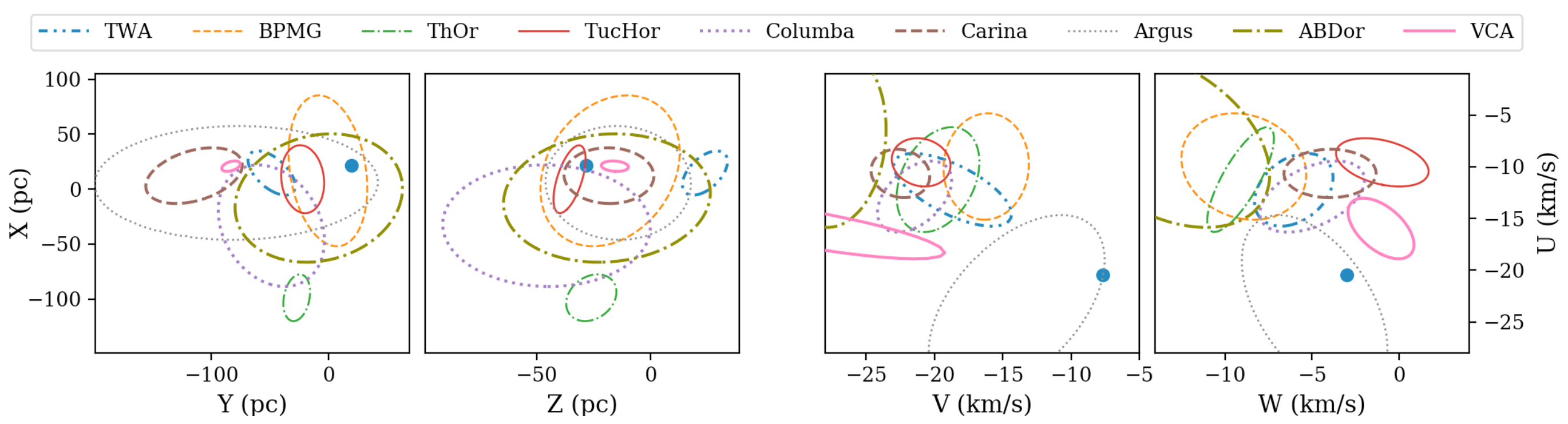}
    \caption{XYZUVW position of HD~206893 (blue point) compared to moving group models from \citet{lee19}.}
    \label{fig:leesong_ymg}
\end{figure*}

The existence and age of the Argus association have been subject of debate, with estimates ranging from $\sim$40~Myr \citep{torres08} to 268~Myr \citep[][]{bell15}. The reality of the association has been called into question \citep{bell15}, however, it has recently been re-established as a field association of age 40--50~Myr by \citet{zuckerman19}. From the low BANYAN~$\Sigma$ probabilities of both Argus and field membership ($<80-90$\%), no definitive conclusion can be drawn. However, due to the unique kinematics of Argus and signatures of youth among its member stars, the association is also included in a recent Bayesian analysis of nearby young moving groups by \citet{lee19}. Calculating the membership probabilities using the models of \citet{lee19} yields similar results to BANYAN~$\Sigma$, with p(Argus) = 63\%, p(Field) = 37\%; kinematic comparisons are shown in Figure~\ref{fig:leesong_ymg}. Given the slight preference for Argus membership from both moving group assessments, we consider a range of possible ages in estimating companion properties. Specifically, we use the \citet{baraffe02} DUSTY and \citet{baraffe03} COND models at 50, 100, 120, and 500 Myr to estimate companion mass in Section~\ref{sec:spec-phot}, where the upper limit on age is set by the stellar SED analysis. At the same time, we note that the significant MIR and FIR excess of the system \citep{chen14} also likely points to a younger age.

\begin{figure*}[h]
    \centering
    \includegraphics[width=0.7\textwidth]{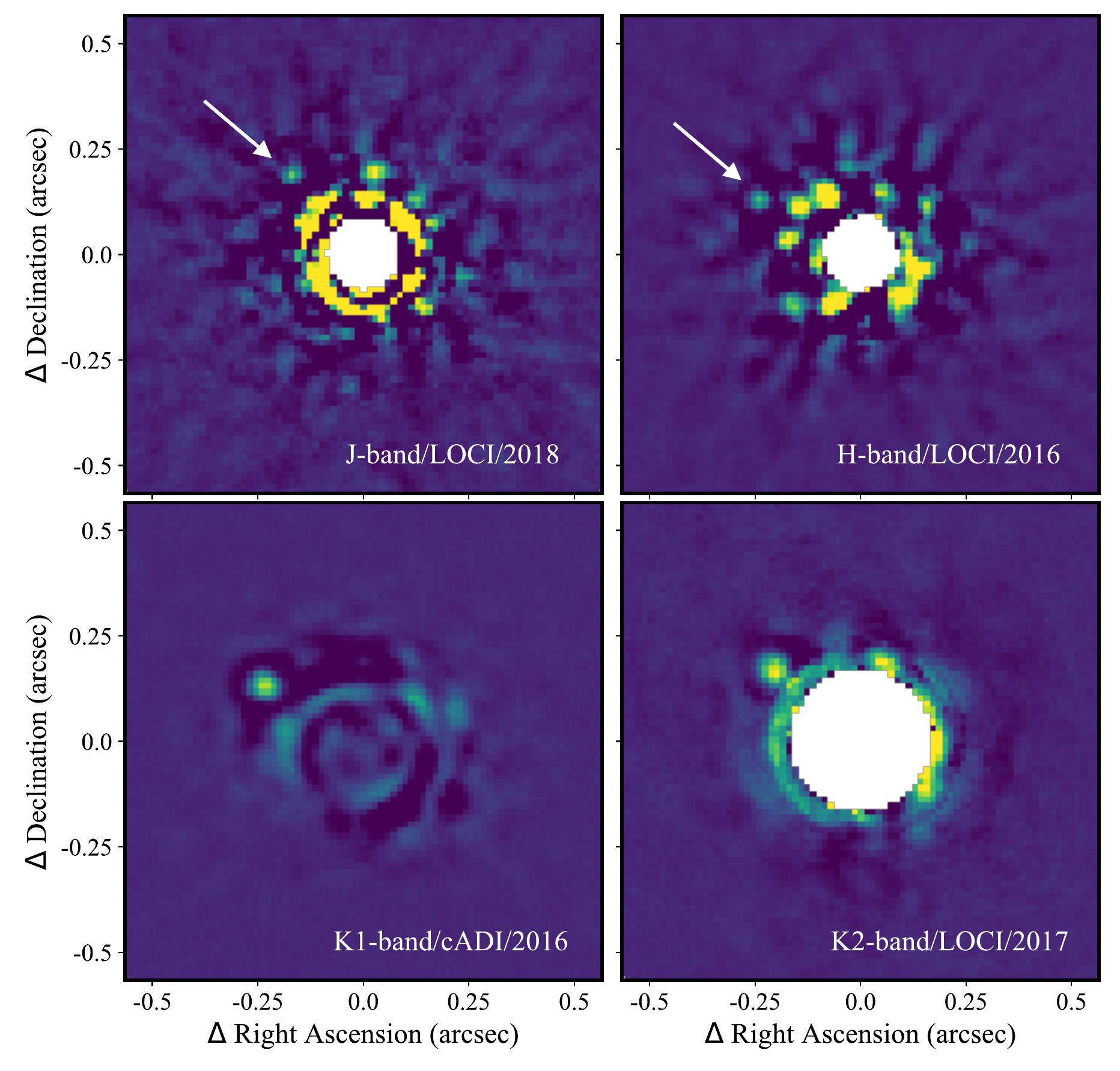}
    \caption{GPI PSF-subtracted images of HD~206893~B in $J$-band (upper left), $H$-band (upper right), $K1$-band (lower left), and $K2$-band (lower right), from the LOCI and cADI reductions. Each individual image is median-collapsed and scaled separately to minimize the contributions of residual speckle noise. Noticeable change in the companion position angle can be seen between the earliest 2016 epoch (upper right) and the latest 2018 epoch (upper left).}
    \label{fig:cadigallery}
\end{figure*}

\clearpage


\section{Results} 
\label{sec:results}

\subsection{GPI Images and Integrated Photometry}
\label{sec:spec-phot}
Figure~\ref{fig:cadigallery} shows the final median-collapsed GPI images in $J$, $H$, $K1$ and $K2$ bands from the LOCI and cADI reductions. The images show significant orbital motion of the companion between the earliest 2016 $H$ epoch and the latest 2018 $J$ epoch, as well as the higher SNR detections at redder wavelengths, consistent with the upward slope of the observed spectrum. 



\begin{table*}[htbp]
\centering
\begin{tabular}{lcccc}
\hline
\hline
System Property       & \multicolumn{2}{c}{Value}                                & Unit        & Ref.  \\
\hline
Parallax       & \multicolumn{2}{c}{$26.08 \pm 0.53$, $24.51 \pm 0.06$}   & mas         & 1, 2  \\
Distance       & \multicolumn{2}{c}{$38.3 \pm 0.8$, $40.8 \pm 0.1$}       & pc          & 1, 2  \\
$\mu_{\alpha}$ & \multicolumn{2}{c}{$93.67 \pm 0.66$, $93.78 \pm 0.09$}   & mas/yr      & 1, 2  \\
$\mu_{\delta}$ & \multicolumn{2}{c}{$0.33 \pm 0.37$, $0.02 \pm 0.08$}     & mas/yr      & 1, 2  \\
Radial Velocity & \multicolumn{2}{c}{$-12.45 \pm 0.59$}                     & km/s        & 2  \\
Age            & \multicolumn{2}{c}{$\sim$50 (kinematics); 601$^{+420}_{-380}$ (stellar SED)}                               & Myr         & 1, 3, 4     \\
$A_{V}$        & \multicolumn{2}{c}{0.07$^{+0.05}_{-0.04}$}                                  & mag         & 4 \\
\hline
\hline
               & HD~206893             & HD~206893~B                       &                            &       \\
\hline               
Spectral Type  & F5V                        & L4 -- L8$^{\dagger}$                                   & -                  & 1 / 4 \\
$M_{J}$        & $2.82 \pm 0.06$            & $15.39 \pm 0.04$                           & mag                        & 4     \\
$M_{H}$        & $2.64 \pm 0.06$            & $13.68 \pm 0.04$                            & mag                        & 4     \\
$M_{Ks}$       & $2.54 \pm 0.06$            & $12.00 \pm 0.07$                            & mag                        & 4     \\
$M_{L'}$       & 2.48                       & $10.39 \pm 0.22$                            & mag                        & 1 / 4 \\
$J$            & $5.87 \pm 0.02$            & $18.44 \pm 0.03$                           & mag                        & 4     \\
$H$            & $5.69 \pm 0.03$            & $16.73 \pm 0.03$                           & mag                        & 4     \\
$Ks$           & $5.59 \pm 0.02$            & $15.05 \pm 0.07$                            & mag                        & 4     \\
$J_\textrm{MKO}$      &                            & $18.38 \pm 0.03$                            & mag                        & 4     \\
$H_\textrm{MKO}$      &                            & $16.82 \pm 0.03$                            & mag                        & 4     \\
$K_\textrm{MKO}$      &                            & $15.02 \pm 0.07$                            & mag                        & 4     \\
$L^{'}$        & 5.52                       & $13.43^{+0.17}_{-0.15}$                     & mag                        & 1 / 3 \\
Mass           & $1.31 \pm 0.03$ & 12 / 20 / 20 / 40 & M$_{\odot}$, M$_\textnormal{Jup}$                          & 4     \\
T$_\textrm{eff}$           & $6500\pm100^{*}$ & 1200--1800$^{\dagger}$ & K                          & 3,4     \\
log($g$)                   & $4.45\pm0.15^{*}$ & 3.0--5.0$^{\dagger}$ & $\log_{10}$(cm/s$^{2}$)      & 3,4     \\
\hline               
\end{tabular}
\caption{Properties of the HD~206893 system. Photometry is provided in the 2MASS system, with MKO also provided for the companion. Companion photometry has been synthesized by integrating the GPI spectra over the relevant bandpasses accounting for differences in the 2MASS and MKO transmission profiles, as described in Section~\ref{sec:spec-phot}. Estimated masses for the companion in $M_\textrm{Jup}$ are provided assuming a range of ages at 50, 100, 120, and 500~Myr. $^{*}$Stellar parameters derived from FEROS spectroscopy by \cite{delorme17}. $^{\dagger}$The large reported ranges in spectral type, effective temperature, and surface gravity reflect the difficulty in matching the observed companion spectrum to L-dwarf standard templates and atmospheric models; see Sections~\ref{sec:results_spec} and \ref{sec:modeling}. \textbf{References.} (1) \citet{milli17}, (2) \textit{Gaia} DR2, \citet{gaiadr2}, (3) \citet{delorme17}, (4) This work.  \label{tab:starcomp_properties}} 
\end{table*}



\begin{figure*}[h]
    \centering
    \includegraphics[width=0.99\textwidth]{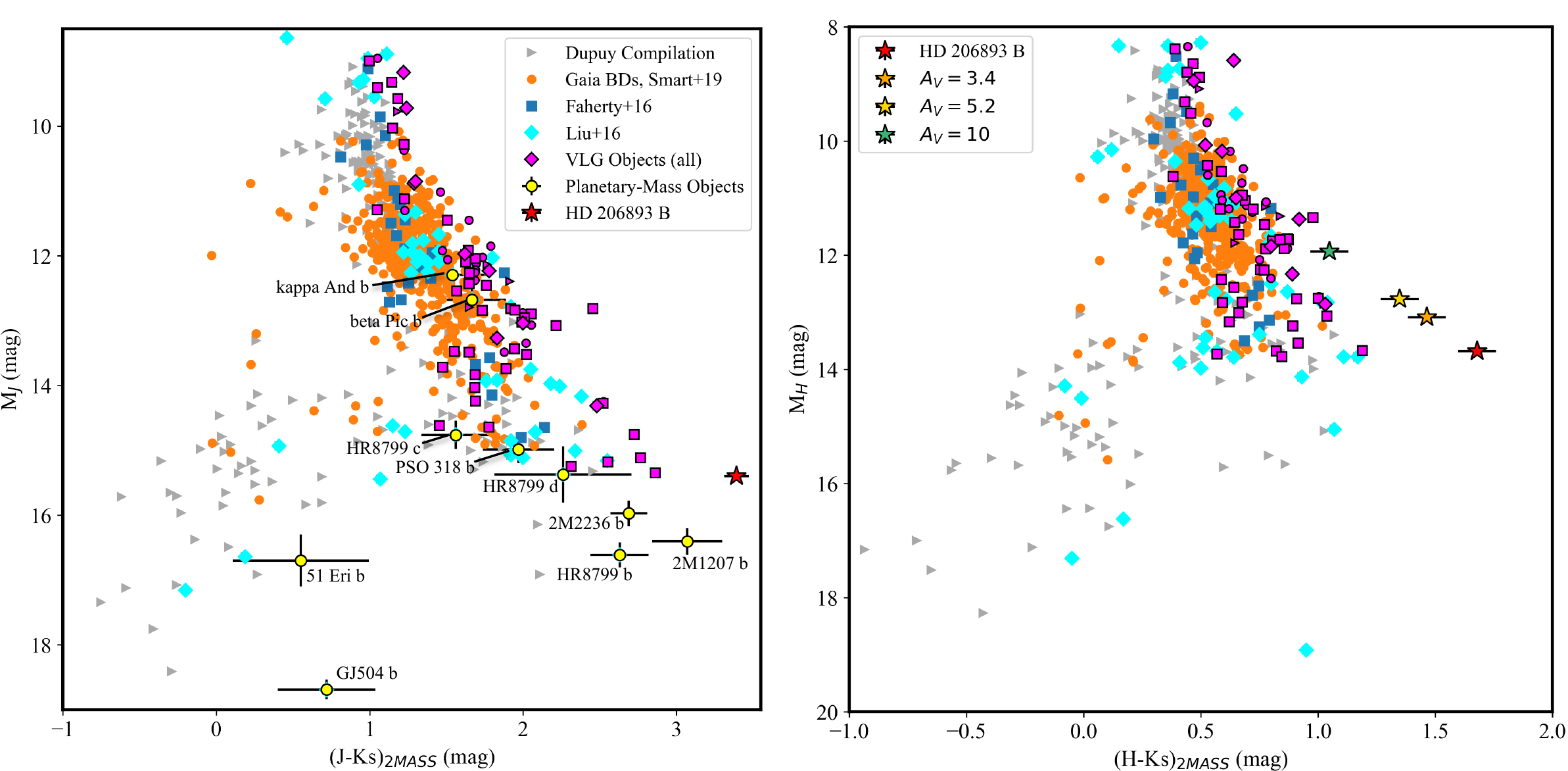}
    \caption{NIR color-magnitude diagrams of L- and T-type brown dwarfs and known planetary-mass objects in the 2MASS photometric system. Data are compiled from the following resources: field brown dwarfs with \emph{Gaia} parallaxes from \citet{smart19} (orange circles); the Database of Ultracool Parallaxes (gray triangles; \url{http://www.as.utexas.edu/~tdupuy/plx/Database_of_Ultracool_Parallaxes.html}) maintained by T. Dupuy, originally in \citet{dupuy12}, \citet{dupuykraus13}, and \citet{liu16} (shown separately in cyan diamonds); and substellar objects considered analogous to planets (blue squares) from \citet{faherty16}. Objects with low-gravity indicators are shown in magenta, demonstrating the known red offset of these objects from the field sequence. The left panel shows the $M_{J}$ vs. $J-Ks$ CMD, with HD~206893~B indicated by a red star and benchmark planetary mass objects labeled (yellow circles). The right panel shows the $M_{H}$ vs. $H-Ks$ CMD. At $H-Ks = 1.68$, HD~206893~B occupies a far redder space in the CMD than any other known substellar object. The additional points for HD~206893~B correspond to $A_{V}$ values of 10 to 3.4 (left to right), adopted to match a similar range of extinction values estimated for the dusty, younger CT Cha system, and $A_{V}=10$ shown in the green star to indicate the extreme values needed to reconcile the object position with the red edge of the substellar sequence.    \label{fig:cmd}}
\end{figure*}

Broadband photometry for HD~206893~B was estimated from the GPI spectral datacubes by interpolating both 2MASS and Mauna Kea Observatory (MKO) transmission filter profiles \citep{tokunaga02} to the GPI wavelength scaling. The $J$, $H$, and $Ks$ (or $K$) filter profiles were convolved with a Gaussian filter of FWHM corresponding to each GPI band, then interpolated to the corresponding GPI wavelength grid. The GPI spectra (combining both $K1$ and $K2$ as described in Section~\ref{sec:reduce_spec} to cover the full $K$ band) were then integrated over the $J$, $H$, and $Ks$ or $K$ bandpasses to derive the magnitudes in Table~\ref{tab:starcomp_properties}, accounting for zero point corrections. The derived photometry allows for positioning HD~206893~B on NIR color-magnitude diagrams (CMDs) of substellar objects, as shown in Figure~\ref{fig:cmd}. The derived photometry from GPI is consistent within uncertainties with the VLT/SPHERE magnitudes reported in \citet{milli17} and \cite{delorme17}. The position of HD~206893~B on the NIR CMDs demonstrates its extraordinary red color, distinct among even the reddest planetary-mass companions, free-floating objects, and low-gravity and/or otherwise peculiar field objects.

\clearpage 

\subsection{Composite Spectrum and Spectral Classification}
\label{sec:results_spec}
\begin{figure*}[h]
    \centering
    \includegraphics[width=0.8\textwidth]{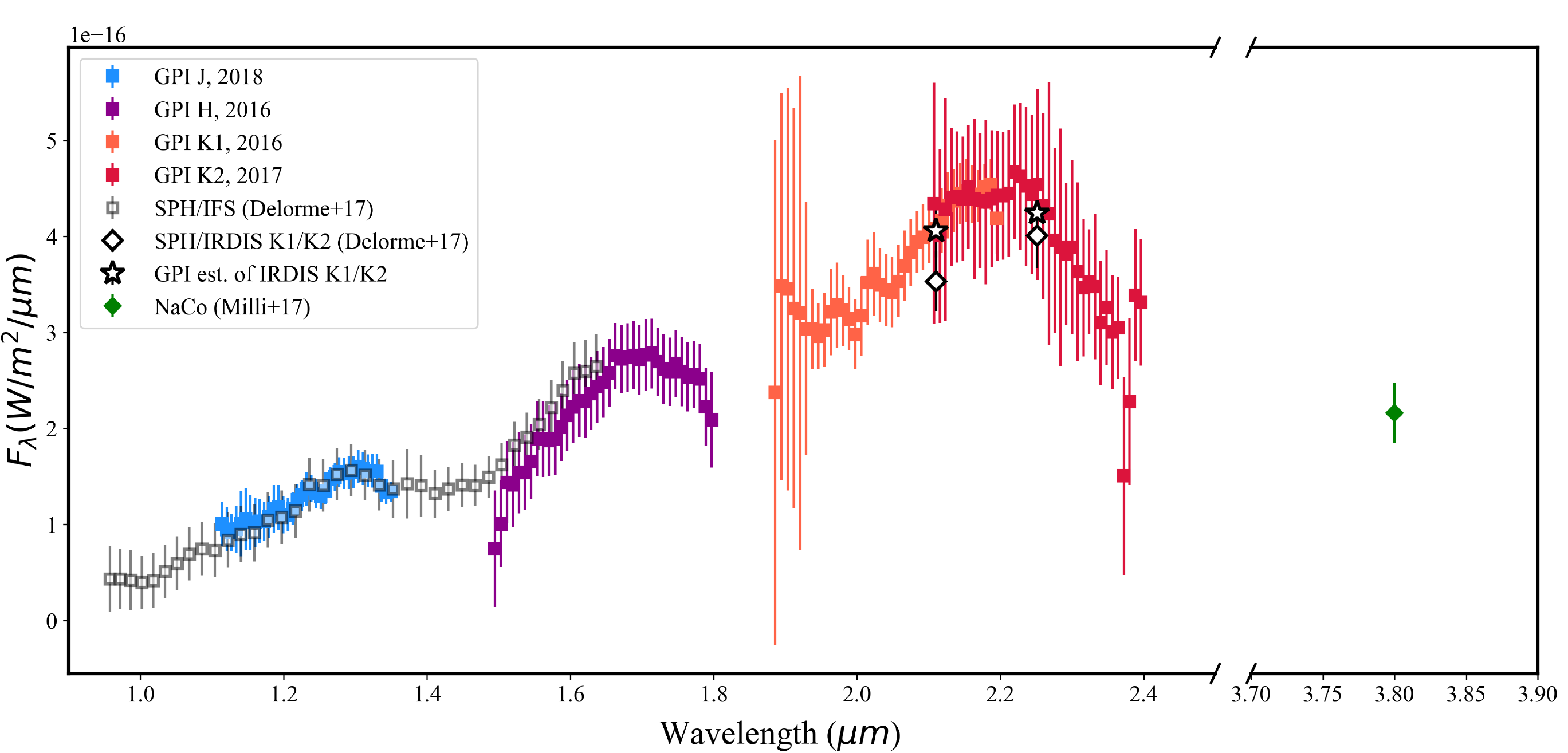}
    \caption{Combined GPI $J$, $H$, $K1$, and $K2$ spectra of HD~206893~B, overlaid with the spectra and photometry from \citet{milli17} and \citet{delorme17}, which show good agreement between the SPHERE IFS and GPI spectroscopy. IRDIS $K1/K2$ photometry has been synthesized from the GPI spectra (white stars) to enable a comparison with the measured SPHERE photometry (white diamonds).}
    \label{fig:spectrum}
\end{figure*}

The resulting combined spectrum is shown in Figure~\ref{fig:spectrum} with each of the extractions from the reduced, post-processed $J$, $H$, $K1$, and $K2$ GPI datacubes. The VLT spectra and photometry from \citet{milli17} and \citet{delorme17} are also shown, demonstrating good agreement between GPI and SPHERE. The GPI $K2$ spectral slope beyond \SI{2.2}{\um} was noted to be unusually steep in the first epoch of observations, and motivated a deeper second-epoch set of observations. The spectral morphology persisted between the two epochs of $K2$ data taken two years apart, and the steepness appears to be dependent upon the spectral extraction as described in Appendix~\ref{sec:appendix_pyklip_covar}. For the remaining analysis in this work, only the extraction from the higher SNR $K2$ spectrum is used.

To determine the spectral properties of HD~206893~B in comparison with known brown dwarfs and atmospheric models, the combined spectra from SPHERE and GPI spanning $Y$ to $K2$ were analyzed using the SpeX Prism Library Analysis Toolkit \citep[SPLAT;][]{burgasser17}. Spectral typing was performed within SPLAT with a $\chi^{2}$ minimization approach, classifying the object by template across the full ensemble of L0--T5 spectra in the SpeX Prism Library. Based on the exceptional redness of the spectrum, the resulting matches within the spectral library correspond to those of the coolest L-type objects, at L6$\pm$1 subclasses. While no single spectrum provides a close morphological match across the full $Y$--$K2$ range, the three closest matches within the spectral library were late-type, low-gravity objects, all with reduced $\chi^2$ values of $\sim$9: WISE~J174102.78-464225.5 \citep[$>$L5$\gamma$;][]{faherty16}, 2MASS~J11193254-1137466 \citep[L7$\gamma$, likely TW~Hya member;][]{kellogg15}, and 2MASS~J11472421-2040204 \citep[L7$\gamma$, likely TW~Hya member;][]{kellogg16}. When the SPLAT library comparison is extended to earlier spectral types, the closest matching objects correspond to M9 pre-main sequence stars with extreme levels of extinction (e.g., 2MASS~J03444520+3201197 in IC~348 and 2MASS~J04325026+2422115 in Taurus, with reduced $\chi^2$ values of 7--8); given the distance and low extinction to HD~206893 and its companion absolute magnitude, these matches are not likely to share similar physical properties to HD~206893~B, and goodness-of-fit is driven by the redness of the object alone.

\begin{figure*}
    \centering
    \includegraphics[width=0.75\textwidth]{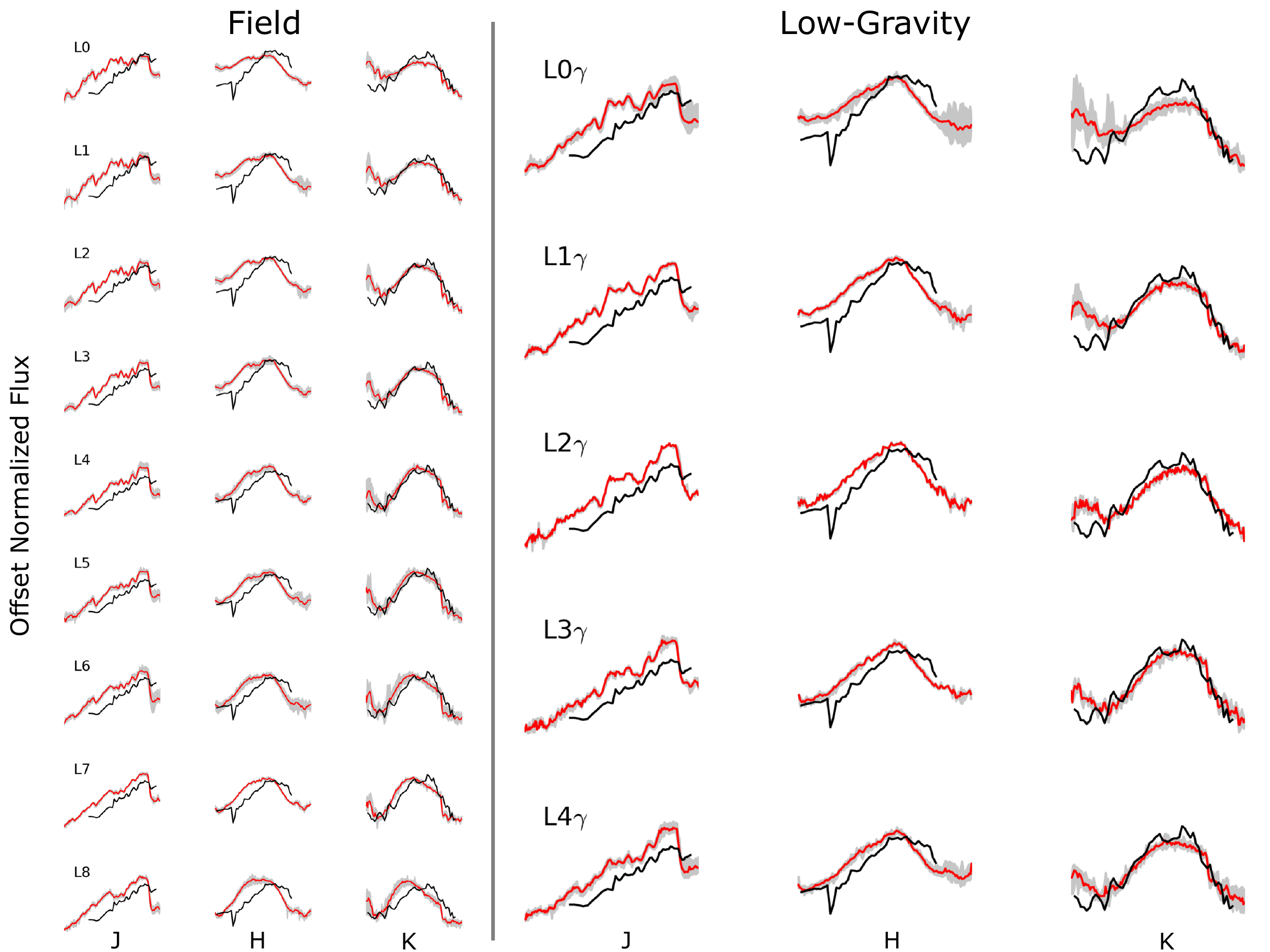}
    \caption{Comparison of the combined GPI+SPHERE spectra (black) to a suite of L-dwarf standard templates of varying gravity (red) using the Ultracool Typing Kit \citep{UTK}. The low-gravity ($\gamma$) standards from \citet{cruz18} include only spectral types up to L4. Each band is independently normalized using the band-specific normalization ranges described in \citet{cruz18}: 0.87-1.39\si{\um} for \textit{zJ}, 1.41-1.89\si{\um} for \textit{H}, and 1.91-2.39\si{\um} for \textit{K}. Morphologically, the HD~206893~B spectrum appears most similar to late, low-gravity objects.}
    \label{fig:UTK}
\end{figure*}

\begin{figure*}
    \centering
    \includegraphics[width=0.95\textwidth]{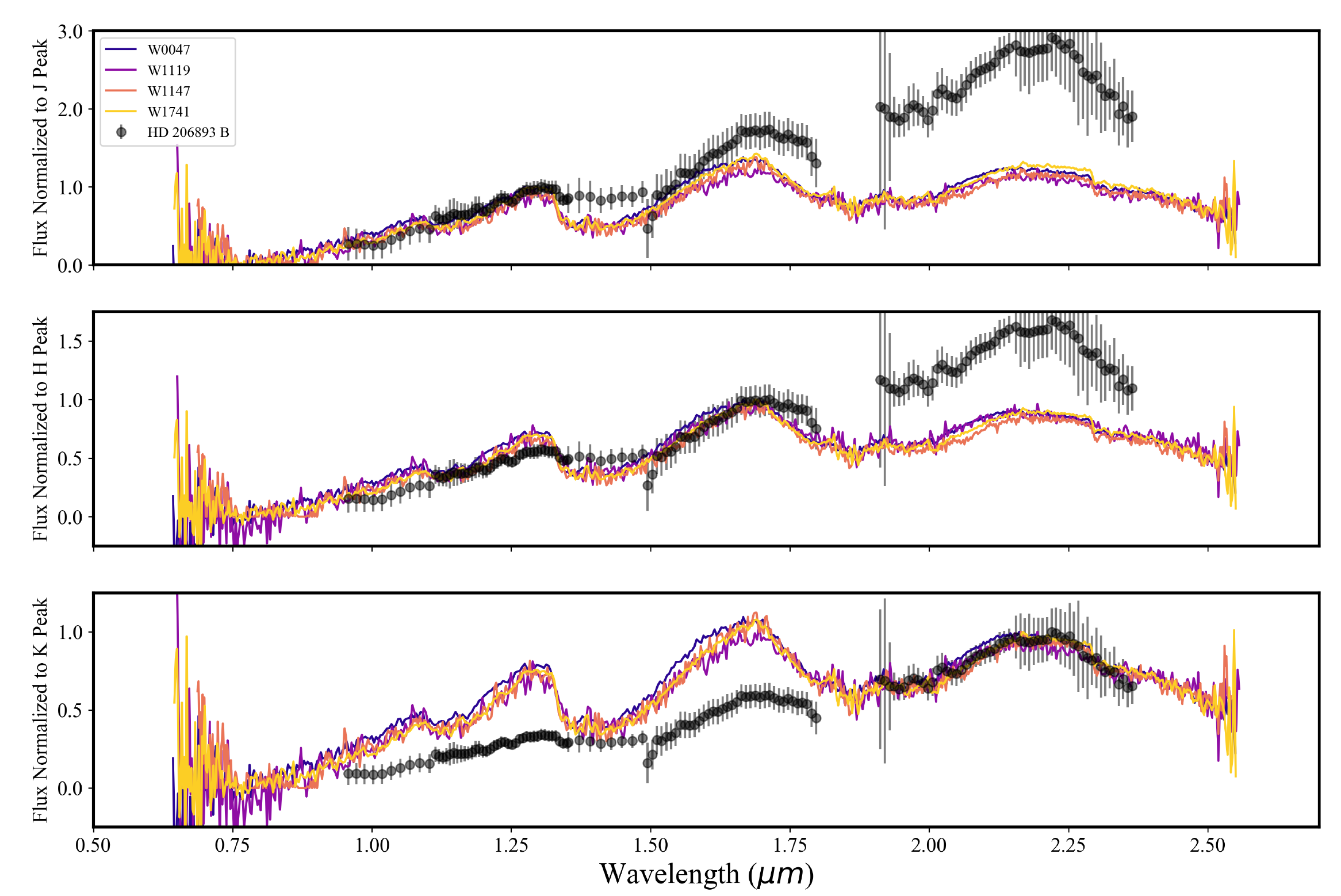}
    \caption{Comparison of the combined GPI+SPHERE spectra (black) to young, low-gravity L5--L7 objects, including the closest substellar spectral matches in the SpeX prism library. Each panel has been normalized on a band-by-band basis to the peak of $J$, $H$, or $K$ (top to bottom). Normalization at shorter wavelengths shows good morphological matches to each of $J$ and $H$ and emphasizes the shallowness of the water band at \SI{1.4}{\um}.}
    \label{fig:lateLs}
\end{figure*}

The large observed spread in NIR colors observed among L-dwarfs of the same optical spectral type \citep[e.g.,][]{leggett03, faherty13} has motivated efforts to produce a systematic means of spectral typing, including qualitative comparison on a band-by-band basis to account for variations in NIR spectral morphology for objects of the same optical spectral classification. Following the work of \citet{cruz18}, which aims to reconcile differences in NIR spectral features for the L-dwarf population, we compare the HD~206893~B spectrum to an ensemble of L-dwarf field and low-gravity standards using the Ultracool Typing Kit \citep[UTK\footnote{\url{https://github.com/BDNYC/UltracoolTypingKit}};][]{UTK}. The UTK analysis is shown in Figure~\ref{fig:UTK}, and involves normalizing each $J$, $H$, or $K$ spectrum individually and comparing the normalized spectra with the spectral templates. The comparison demonstrates the dissimmilarities between HD~206893~B and typical L-dwarfs. The spectra are qualitatively similar to the later field and young L populations, commensurate with the SPLAT typing analysis. However, the steep slope of the blue side of $H$-band and peaked morphology more closely match the low-gravity population than field objects, potentially indicating youth \citep[although challenges exist in using the triangular shape to reliably distinguish dusty atmospheres from low-gravity ones;][]{allersliu13}. Figure~\ref{fig:lateLs} shows the three closest spectral matches from SPLAT, including 10~Myr old candidate TW~Hya members, in addition to the unusually red AB~Doradus member WISEP~J004701.06+680352.1 \citep{gizis12}. Spectra have been normalized per band over the same normalization ranges used in UTK, extending a simliar analysis to later low-gravity L-dwarfs. These late-type comparisons show the greatest morphological similarities to HD~206893~B, particularly across $J$ and $H$ band when normalized on a band-by-band basis. As such, we adopt a conservative range of L6$\pm$2 for the companion spectral type, noting stronger similarities to young low-gravity objects in this spectral range than to the late-type field population seen in Figure~\ref{fig:UTK}.

\clearpage

\begin{table}[htbp]
\ssmall
\begin{tabular}{lccl}
\hline
\hline

\textbf{Model Grid} & \textbf{T$_\textnormal{eff}$ (K)} & \textbf{log[$g$ (cgs)]} & \textbf{Notes}   \\
\hline
Sonora              & 1100-1600         & 3.75-5.0   & Solar metallicity, f$_\textnormal{sed}$ = 1    \\

Cloudy-AE 60        & 600-1700          & 3.5-5.0    & Solar metallicity, fixed \SI{60}{\um} particles  \\

DRIFT-PHOENIX       & 1000-2000         & 3.0-5.0    & Solar metallicity, super-solar {[}M/H{]} = 0.3 \\

BT-Settl            & 1200-2050         & 2.5-5.5    & Solar metallicity                                           \\
\hline
\end{tabular}
\caption{\label{tab:modelparams} Parameter ranges in the model grid search. For all models tested, the object radius was scaled by factors ranging from 0.5 - 3.5 R$_\textnormal{Jup}$ in the model fitting, in steps of 0.01.}

\end{table}

\subsection{Brown Dwarf Atmospheric Modeling}
\label{sec:modeling}

\begin{figure*}
    \centering
    \includegraphics[width=1.0\textwidth]{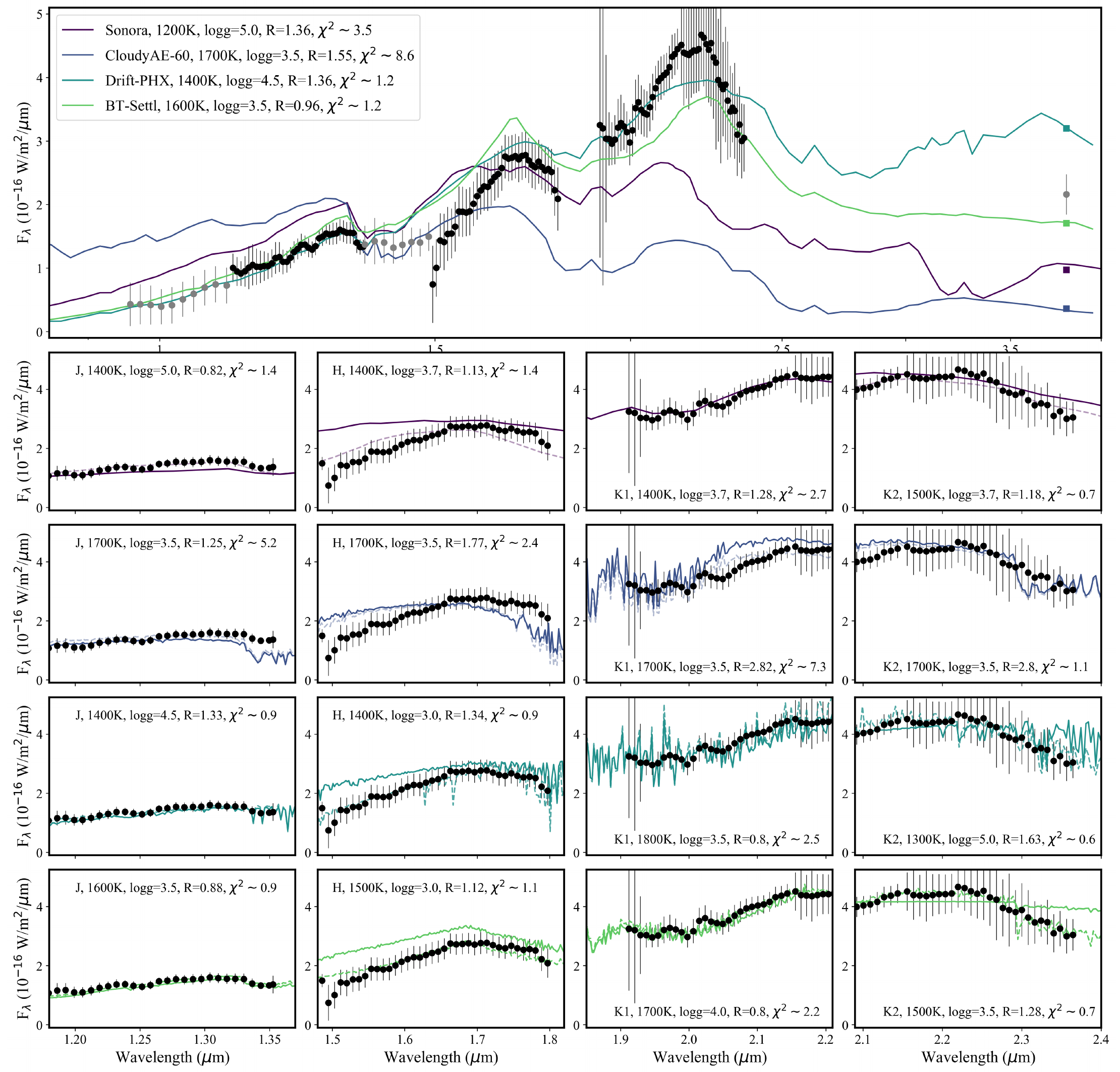}
    \caption{Best-fit models from a grid comparison of the GPI spectra (black points) and SPHERE+NaCo spectra and photometry (gray points) with four theoretical atmosphere models (Sonora in purple--Marley et al. 2020 in prep.; Cloudy-AE60 in blue --\citet{madhu11}; DRIFT-PHOENIX in dark green--\citet{helling08, witte09, witte11}; and BT-Settl in light green--\citet{allard12}). The topmost panel shows the best-fits to the full spectral coverage from the combined SPHERE, GPI, and NaCo data. The integrated $L'$ flux is shown for each set of models in square symbols. The four rows of subpanels show the results from fitting the four GPI bands separately. Corresponding $\chi^2$ best-fit model parameters with (solid lines) and without (faint dashed lines) incorporation of covariances are shown in each panel.}
    \label{fig:all_best_models}
\end{figure*}

The combined spectra and photometry of HD~206893~B from GPI, SPHERE, and NaCo, spanning $0.95$ to $3.8\mu$m, were compared to four grids of theoretical atmosphere models in order to derive estimates of effective temperature, surface gravity, and metallicity. The model grids used were: \textbf{(1)} a subset of the Sonora models, a new model grid designed for substellar and young giant planet atmospheres \citep[Marley et al. 2020, in prep.; see also the currently available solar metallicity cloud-free grid from][]{marley2018}\footnote{\url{http://doi.org/10.5281/zenodo.1309035}}, \textbf{(2)} the Cloudy-AE\footnote{\url{https://www.astro.princeton.edu/~burrows/8799/8799.html}} models, originally developed for the HR~8799 planets \citep[][]{madhu11}, \textbf{(3)} the DRIFT-PHOENIX\footnote{\url{http://svo2.cab.inta-csic.es/theory/newov}} Atmosphere Models \citep{helling08, witte09, witte11}, and \textbf{(4)} the BT-Settl (2015)\footnote{\url{https://phoenix.ens-lyon.fr/Grids/BT-Settl/CIFIST20112015}} grid \citep{allard12, allard14}. For each grid, individual models spanning ranges of $T_\textnormal{eff}$ and log($g$) space were binned to the resolution of the GPI spectra and interpolated over the GPI wavelength axes. All models used were of solar metallicity, with the exception of DRIFT-PHOENIX, which included both solar and super-solar $[M/H]$ values. Object radius was treated as a fit parameter, with each model scaled to match the object SED using the \textit{Gaia}~DR2 distance and a range of plausible object radii. The explored parameter ranges for the model grids are summarized in Table~\ref{tab:modelparams}, scaling the object radius from 0.5--3.5~R$_\textnormal{Jup}$ in steps of 0.01. The resulting $\chi^{2}$ fit statistics were calculated accounting for covariances in spectral channels, following the approach outlined in \citet{derosa16}, with zero-valued covariance assigned to the SPHERE IFS $Y$ and $J/H$ data not covered by GPI. Best fit models were determined separately for each band and for the full spectrum, allowing us to investigate disparities between best fit models between wavelength regimes. The resulting best fit models from the Sonora, DRIFT-PHOENIX, Cloudy-AE, and BT-Settl grids are shown in Figure~\ref{fig:all_best_models} and incorporate covariances. Results from a standard covariance-free $\chi^{2}$ approach are also shown on a band-by-band basis with faint dashed lines in Figure~\ref{fig:all_best_models}, the results of which are summarized in Appendix~\ref{sec:appendix_models}. While the standard fits without covariance estimation sometimes more closely visually match the GPI spectral data points, the derived physical parameters are in better agreement across bands when accounting for spectral covariance.

Given the extraordinary redness of the companion, none of the models provides a qualitatively good fit to the full spectrum, and reduced $\chi^{2}$ statistic values range from $\sim$1.2~to~8. The best fits to the Sonora and Cloudy-AE60 models all lie at the edge of the available grid parameters (Table~\ref{tab:modelparams}), suggesting that the best-fits likely reside outside of the range of the current model grids. As noted in \citet{delorme17}, the unusual SED of the object and its extreme spectral features present significant discrepancies relative to existing model atmosphere grids, necessitating additional sources of reddening. Challenges in atmospheric model fitting of L-dwarfs, particularly in the reproduction of the NIR spectral slope and H-band morphology of young/low-gravity objects, have been noted in previous studies \citep[e.g.,][]{manjavacas14} and have been attributed to the treatment of dust in current theoretical atmosphere models. We explore this possibility further in Section~\ref{sec:reddening}, and note that determining the cause of mismatches between observations and current models may involve considering a wide range of particle properties, overall dust content, and multispecies gas opacities, tasks well-suited for retrieval-based methods.

When we fit the full spectrum, the best-fit results from each model grid show a range of physical parameters; temperatures range from 1200 to 1700~K, surface gravities from log($g$) of 3.5~to~5.0, and object radii from 0.96~to~1.55~R$_\textnormal{Jup}$. However, the derived properties vary significantly when comparing results across fits to individual bands. The derived log($g$) values from the model fits and the inferred radii necessary to scale the flux are inconsistent, a discrepancy noted for previous fits of exoplanet spectra and photometry \citep[e.g., HR8799bcd, beta Pic b, 51 Eri b, and others;][]{marois08,marley12,morzinski15,rajan17}. A detailed assessment of the full-spectrum fitting using each model grid, as well as brief summaries of model grid properties, are provided in Appendix~\ref{sec:appendix_models}.

Following the methodology applied for spectral typing the object, we performed fits to the four GPI bands individually for each set of model grids, in addition to the full spectral fit. These fits are shown in the four subpanel rows of Figure~\ref{fig:all_best_models}, and each fit accounts for spectral covariance within that band. Accounting for spectral covariance, the best fit models agree within 100~K across all four bands for the Sonora and CloudyAE-60 models and across $JHK2$ for DRIFT-PHOENIX and BT-Settl. Departures in $K1$ for the DRIFT-PHOENIX and BT-Settl model grids may be attributed to the higher level of spectral covariance in the $K1$ band. Similarly, the high level of spectral correlation observed in $H$-band leads to a mismatch between the best fit models and datapoints for all four suites of models, an effect observed in previous IFS studies \citep[e.g.,][]{rajan17}. For comparison, the best fit models adopting a standard $\chi^{2}$ approach are shown in light dashed lines within each panel. The estimated log($g$) values are consistent across bands only for the Sonora and Cloudy-AE60 models, favoring lower-gravity fits. Each of the other three model suites provide a broad range of surface gravities and radii depending on the band in question.

\subsection{Astrometric Analysis}
\label{sec:astrometry}

The four new astrometric epochs from the GPI observations presented here were combined with five additional astrometric measurements from VLT/SPHERE and VLT/NaCo \citep{milli17, delorme17, grandjean19} to explore the range of potential orbital parameters for HD~206893~B. The full set of astrometric measurements used in this analysis is provided in Table~\ref{tab:astrometry}. Following the astrometric analyses described in \citet{derosa15} and \citet{rameau16}, a parallel-tempered Bayesian MCMC approach using the \textit{emcee} package \citep{foremanmackey13} was used to simultaneously fit eight companion orbital parameters, namely: semi-major axis ($a$), inclination ($i$), eccentricity ($e$), sums and differences of longitude of ascending node and argument of periastron ($\Omega\pm\omega$), epoch of periastron passage ($\tau$), orbital period, and system parallax. We adopt the following standard prior distributions on the orbital parameters: log uniform in $a$, uniform in $e$ and $\cos(i)$, and Gaussian for the system mass and system parallax. We initialized 512 walkers at 16 different temperatures; the lowest temperature samples the posterior distribution, while the highest temperature samples the prior. We advanced each chain for one million steps, saving every hundred steps. The first half of the chain was discarded as a ``burn-in'' period to ensure the final posterior distributions were not a function of the initial positions of the walkers.


\begin{table}[htbp]
\scriptsize
\centering
\begin{tabular}{lcccc}
\hline
\hline

Epoch    & Separation (mas)     & Pos. Angle ($^{\circ}$)                 &  Instrument, Band              &       \\
\hline 
2015-10-05     & $270.4 \pm 2.6$            & $69.95 \pm 0.55$                            & SPH/IRDIS, $H$         & 1     \\
2016-08-08     & $269.0 \pm 10.4$           & $61.6 \pm 1.9$                              & VLT/NaCo, $L'$            & 1     \\
2016-09-16     & $265 \pm 2$                & $62.25 \pm 0.11$                            & SPH/IRDIS, $K1/K2$     & 2     \\
2016-09-22     & $267.6 \pm 2.9$            & $62.72 \pm 0.62$                            & GPI, $H$                  & 3     \\
2016-10-21     & $265.0 \pm 2.7$            & $61.33 \pm 0.64$                            & GPI, $K1$                 & 3     \\
2017-07-14     & $260.3 \pm 2$              & $54.2 \pm 0.4$                              & SPH/IRDIS, $H$         & 4     \\
2017-11-09     & $256.9 \pm 1.1$            & $51.01 \pm 0.35$                            & GPI, $K2$                 & 3     \\
2018-06-20     & $249.11 \pm 1.6$           & $45.5 \pm 0.37$                             & SPH/IRDIS, $H2/H3$     & 4     \\
2018-09-24     & $251.7 \pm 5.4$            & $42.6 \pm 1.6$                              & GPI, $J$                  & 3     \\
\hline               
\end{tabular}
\caption{Astrometry for HD~206893~B. \textbf{References.} (1) \citet{milli17}, (2) \citet{delorme17}, (3) This work, (4) \citet{grandjean19}. \label{tab:astrometry}} 
\end{table}

\begin{figure}
    \centering
    \includegraphics[width=0.48\textwidth]{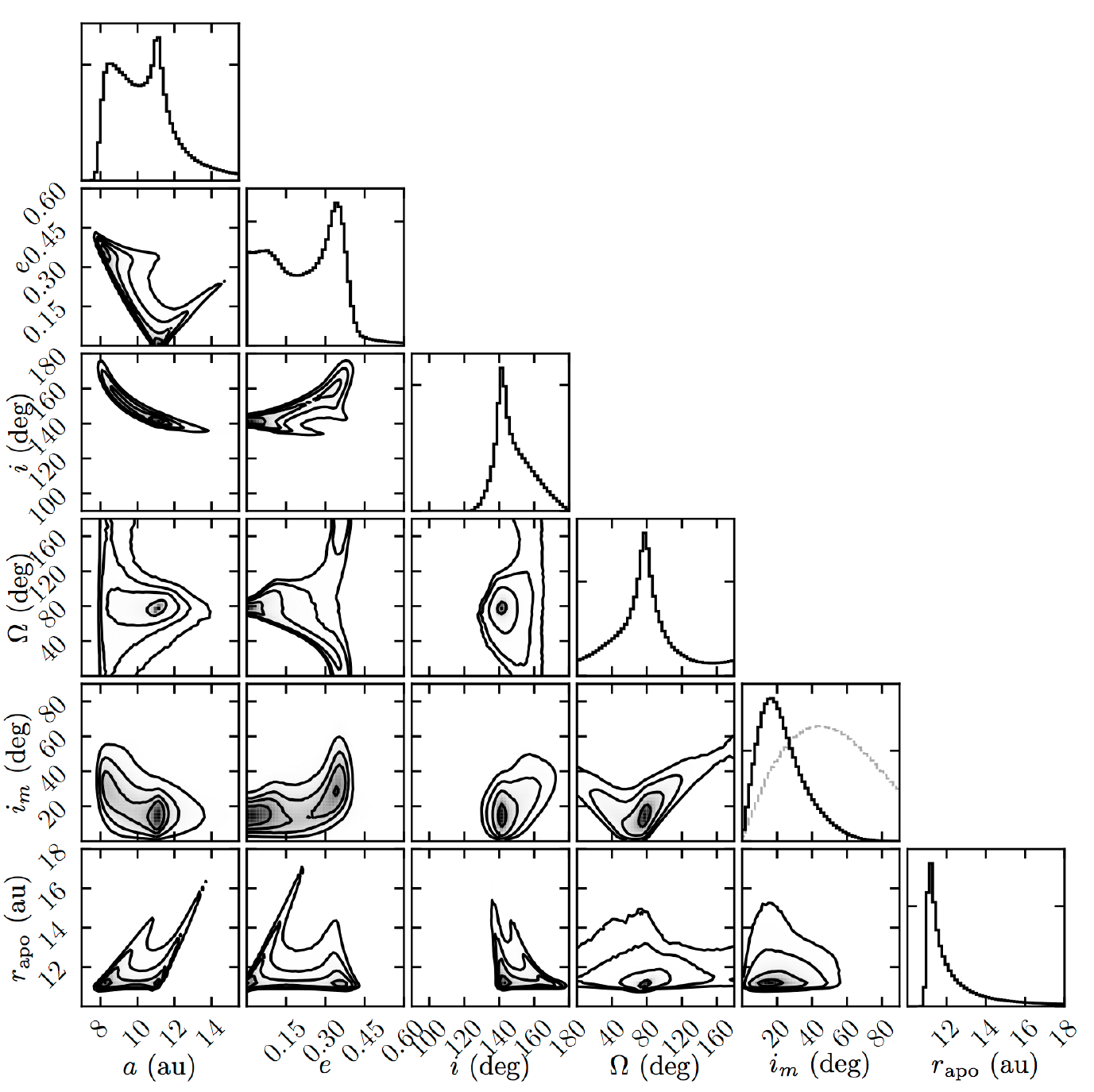}
    \caption{Posterior distributions for semimajor axis, eccentricity, inclination, longitude of the ascending node, mutual inclination of the companion orbit and circumstellar disk \citep[with respect to the inferred disk geometry from][]{milli17}, and the radius of apoapsis. The gray dashed line overplotted represents the prior distribution in mutual inclination, showing that the posterior distribution slightly favors more co-planar over more misaligned configurations.}
    \label{fig:orbit_triangle}
\end{figure}

The resulting posterior distributions for a subset of selected orbital parameters are shown in Figure~\ref{fig:orbit_triangle}. The 1$\sigma$ confidence intervals from the posterior distributions suggest a tightly constrained semi-major axis of $10.4^{+1.8}_{-1.7}$~au, orbital inclination of $145.6^{+13.8}_{-6.6}$\,deg, and apoapsis distance of $11.7^{+2.7}_{-0.6}$~au, and place less restrictive constraints on the eccentricity ($0.23^{+0.13}_{-0.16}$) with a corresponding period range of $29.1^{+8.1}_{-6.7}$~years. The narrower posterior distribution on the apoapse distance is consistent with observing the companion near apastron, with the uncertainty on semimajor axis dominated by a less-stringent constraint on the radius of periapsis. This can also be seen in the covariance of semimajor axis and eccentricity posterior distributions in Figure~\ref{fig:orbit_triangle}, which favor smaller values of $a$ for more eccentric ($e>0.2$) orbits.

We investigated the co-planarity of the orbit of the companion with the spatially-resolved debris disk by comparing the inclination and position angle of the visual orbit on the sky to those estimated for the debris disk by \citet{milli17}. The disk inclination and position angles are defined over more restrictive ranges than for a visual orbit; the disk is not measured to orbit in a particular direction around the star, and a 180 degree ambiguity exists in the position angle measurement. Under the assumption that the disk and orbital plane of the companion are non-orthogonal, we adopt an inclination for the disk of $i_d=140^{\circ}\pm10^{\circ}$, and a position angle of $\Omega_d = 60^{\circ}\pm10^{\circ}$. We drew random variates from two Gaussian distributions describing these parameters for each sample of the posterior distribution of the visual orbit, assuming that the measured geometry of the disk and the fit to the visual orbit are uncorrelated. We measure the mutual inclination, $i_{m}$, between the disk and orbit using the following relation:

\begin{equation}
    \cos(i_m) = \cos(i_c) \cos(i_d) + \sin(i_c) \sin(i_d) \cos(\Omega_c - \Omega_d)
    \label{eqn:mutual_i}
\end{equation}

where the $c$ and $d$ subscripts refer to the companion and disk, respectively. The resulting distribution of $i_m$ and correlations with the other orbital elements are shown in Figure~\ref{fig:orbit_triangle}. The position angle of nodes for the orbit of the companion ($\Omega$) is plotted from 0--180$^{\circ}$, the same restricted range as the position angle of the disk. An equally good fit to the visual orbit can be found with $\omega+\pi$ and $\Omega+\pi$, however these orbits would be significantly misaligned with the assumed disk position angle of $60^{\circ}\pm10^{\circ}$. The distribution of mutual inclinations for two randomly-orientated planes is also shown for reference, and $i_m$ appears significantly shifted towards more co-planar solutions relative to this prior distribution. The current measurement precision for both the disk geometry and the visual orbit are not sufficient to exclude moderately misaligned configurations ($i_{m}\sim20^{\circ}$).

\begin{figure}
    \centering
    \includegraphics[width=0.48\textwidth]{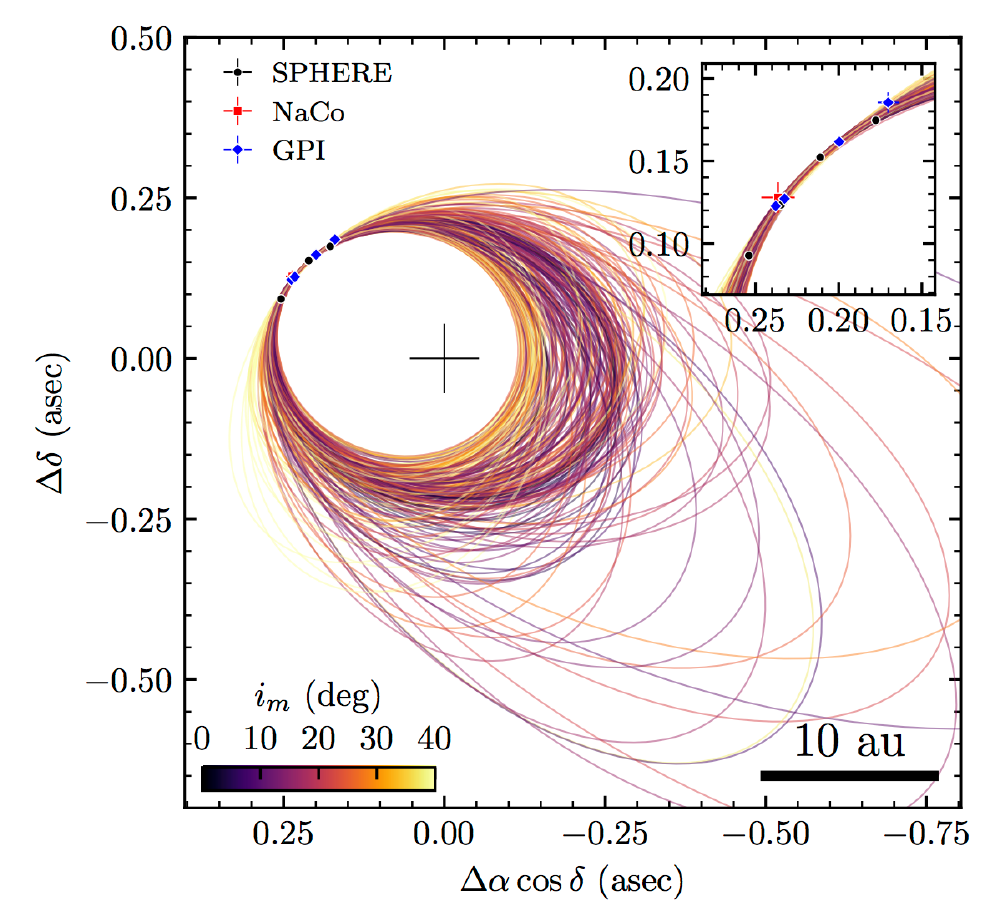}
    \caption{Subsample of 250 randomly-sampled MCMC orbits to HD~206893~B based on four epochs of GPI astrometric monitoring (blue diamonds) in addition to five epochs of VLT/SPHERE (black circles) and VLT/NaCo (red square) astrometry, covering a total time baseline from 2015-2018 (astrometric points and errors shown in inset). The darker orbits correspond to orbits with lower values of mutual inclinations between the companion orbit and the disk plane, with the mutual inclination distribution in Figure~\ref{fig:orbit_triangle} peaking at $\sim20 ^{\circ}$.} 
    \label{fig:orbit}
\end{figure}

A representative sample of 250 orbits drawn randomly from the posterior distribution is shown in Figure~\ref{fig:orbit}. The plotted orbit colors indicate the mutual inclination value for the companion orbit relative to the disk geometry inferred from modeling in \citet{milli17}, favoring more co-planar configurations. Misaligned configurations with greater values of $i_{m}$ are generally more eccentric, corresponding to smaller semimajor axis and periapse distance, which can be seen from the posterior covariances in Figure~\ref{fig:orbit_triangle}. The long tail of the posterior distribution for apoapse distance, $r_\textrm{apo}$, extends to large values coincident with the estimated inner edge of the disk at $\sim50$~au by \citet{milli17}. 


\subsection{Sensitivity to Additional Companions}
\label{sec:completeness}
\begin{figure}[h]
    \centering
    \includegraphics[scale=0.45]{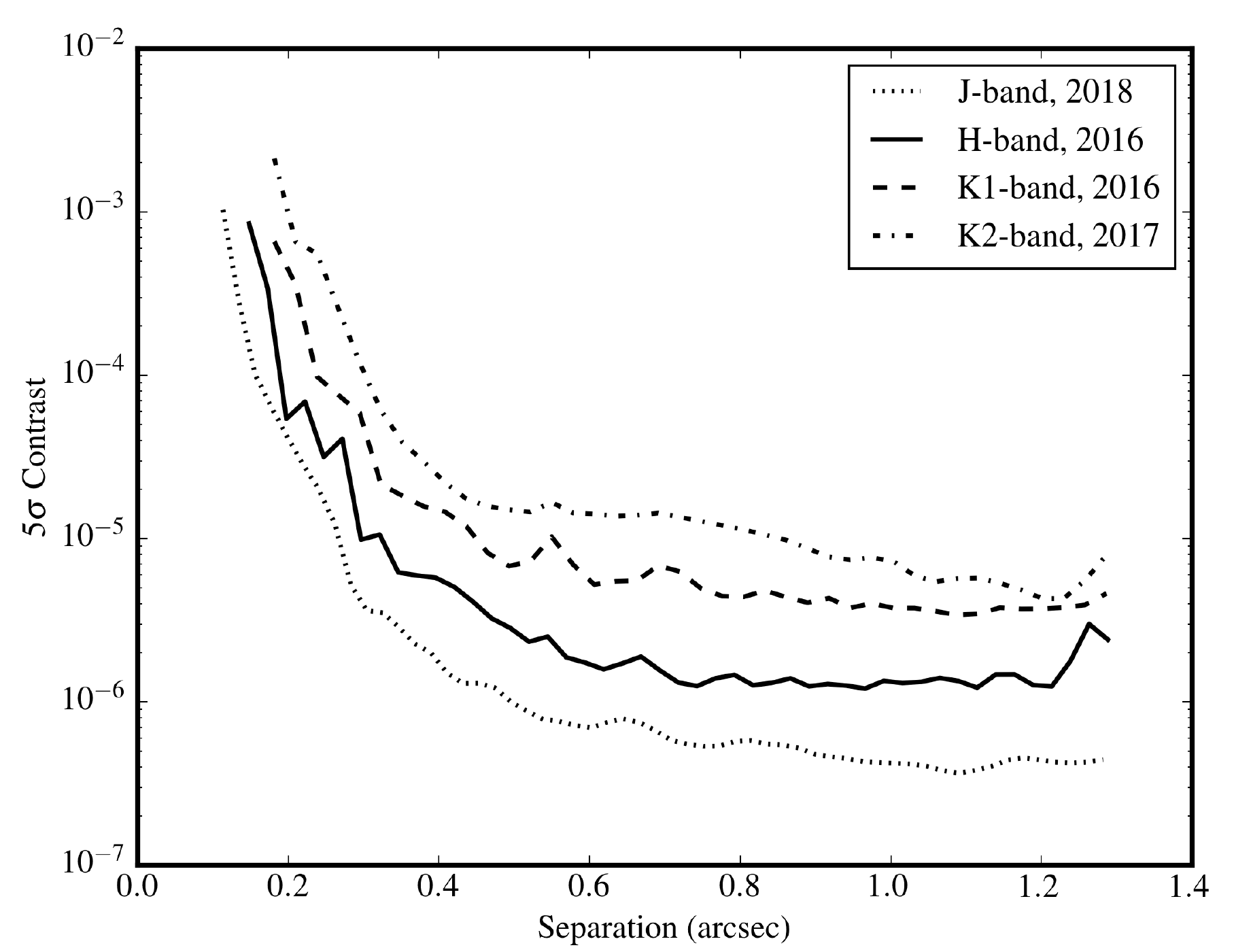}
    \caption{5$\sigma$ contrast curves calculated for the pyKLIP-reduced GPI images in each band of observation, assuming L-type object spectra.}
    \label{fig:contrasts}
\end{figure}

\begin{figure}[h]
    \centering
    \includegraphics[width=0.45\textwidth]{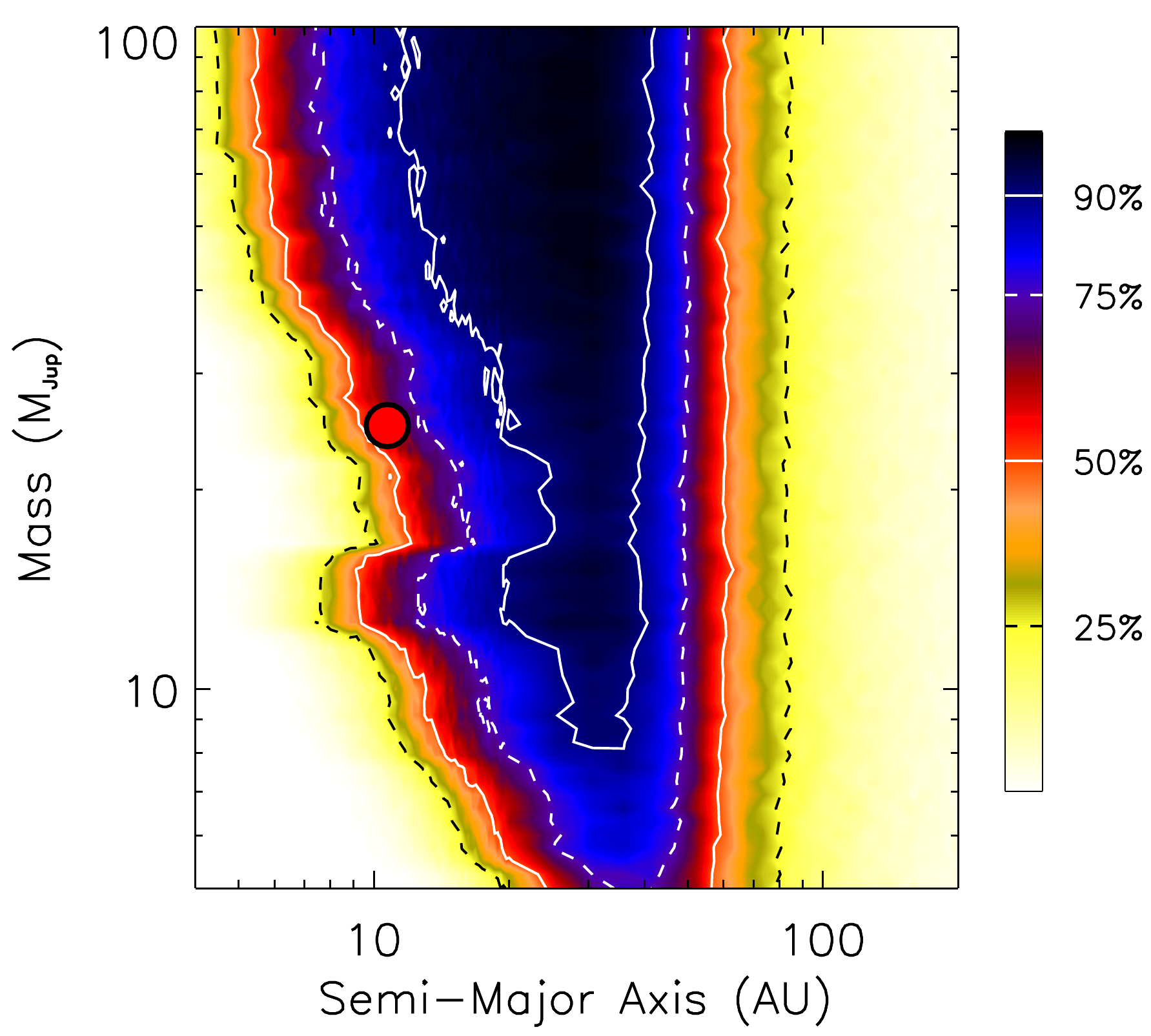}
    \caption{Combined sensitivity of all epochs and bands of GPI observations, expressed in terms of companion masses as a function of orbital separation. PyKLIP contrasts from Figure~\ref{fig:contrasts} are converted into physical parameters following the methodology of \citet{nielsen19} using an assumed age of 250~Myr and the CIFIST2011 BT-Settl atmosphere models \citep{caffau11, allard14, baraffe15} combined with the COND evolutionary sequences \citep{baraffe03}. The `divot' seen at 10~au and $\sim$12~$M_\textnormal{Jup}$ corresponds to the closer inner working angles accessible in only the shorter wavelength observations. The colorbar and scale show corresponding completeness levels, with the approximate mass and semimajor axis of HD~206893~B for reference (red point).}
    \label{fig:completeness}
\end{figure}

From the pyKLIP reductions described in Section~\ref{sec:reduce} and shown in Figure~\ref{fig:gallery}, we calculate contrast curves for each GPI band. These curves are shown in Figure~\ref{fig:contrasts}; they are calculated by estimating noise levels from the PSF-subtracted images and correcting for instrument throughput, calculated by injecting and recovering false planets with L-type spectra \citep[see][]{wang18}. The sensitivity to companions is then estimated from the contrast curves following the methodology of \citet{nielsen19}. In brief, a synthetic population of \num{e4} companions is drawn from a grid sampling mass and semi-major axis values, and each companion is randomly assigned orbital parameters (inclination, eccentricity, argument of periastron, and epoch of periastron passage). Projected separation and magnitude difference are estimated and compared to the contrast curve to determine detection probability. Variation in contrast by epoch is accounted for by stepping the companion orbit forward in time in order to generate a new estimation for companion separation and contrast that can be compared to the observed contrast curve. The recovery rate for companions at a given mass and semimajor axis is then translated into a completeness percentage. The results are shown in Figure~\ref{fig:completeness}, where the mass of HD~206893~B corresponds to an assumed system age of 250~Myr. It is possible to exclude additional companions in the GPI images in the planetary regime down to 5~$M_\textnormal{Jup}$ at $\sim$20~au, and in the brown dwarf regime down to $\sim$10--20~$M_\textnormal{Jup}$ at $\sim$10~au. For comparison, the inner additional companion signal predicted from RV variation by \citet{grandjean19} suggests a $\sim$15~$M_\textnormal{Jup}$ object at an orbital separation of 1.4--2.6~au. The requisite contrast to detect such an object ($<$10$^{-4}$ in the NIR at a separation of 50--60~mas) is beyond the sensitivity limits of current extreme AO direct imaging instruments and can be compared to the contrast curves and sensitivity of our data shown in Figures~\ref{fig:contrasts} and \ref{fig:completeness}.

\clearpage


\section{Discussion} 
\label{sec:discuss}

\subsection{Companion redness and system extinction properties}
Given the existence of circumstellar material in the HD~206893 system, dust extinction may have an effect on the measured companion magnitudes; however, the nature of reddening in the system is currently unknown. The unusual redness of the companion could be explained by a range of reddening sources, namely: conventional interstellar reddening, extinction due to the debris disk, the presence of circumsecondary disk material around the companion itself, or dust within the atmosphere of the companion itself. 

We adopted reddening values of $A_{V} = 3.4$ and $5.2$, in conjunction with the extinction laws of \citet{riekelebofsky85}, to estimate potential shift in color-magnitude space due to extinction caused by dust in the vicinity of the companion. These values for $A_{V}$ were chosen from best-fit extinction estimates for the substellar companion in the CT~Cha disk system \citep{schmidt08, wu15}. As CT~Cha is considerably younger ($\sim$2~Myr) than HD~206893, this is likely an overestimate of potential extinction caused by the far less optically-thick HD~206893 debris disk. The resulting de-reddened CMD positions are shown in the right panel of Figure~\ref{fig:cmd}, with an additional value of $A_{V}=10$ shown for comparison to illustrate the extreme and likely unphysical amount of extinction required to meet the red edge of the field L-dwarf sequence.

To determine which extinction estimates correspond to realistic values for a companion viewed edge-on within a debris disk, we compare dust column density and optical depth estimates to typical circumstellar disk values. Using standard relations for visual extinction and gas column density for the interstellar medium ($N_\textrm{H}$~(\si{\per\square\cm}) = \num{2.21e21} $A_{V}$ (mag); \citealp{guver09}), and translating this relation into a dust column density using the canonical 100:1 gas to dust ratio yields a dust mass column density of \SI{3.7e-4}{\g\per\square\cm} for $A_{V}=10$. For comparison, visual extinction for the edge-on AU~Mic debris disk system with higher fractional infrared luminosity \citep[\num{3.9e-4};][]{matthews15} has been estimated as $A_{V} = 0.5$ \citep{yeejensen10}. Assuming small grains and a typical silicate density (\SI{0.5}{\um} and \SI{3.27}{\g\per\cubic\cm}) yields a particle column density of \SI{2e8}{grains\per\square\cm}, suggesting that a debris disk with $A_{V} = 10$ is unrealistically optically thick ($\tau\sim1.7$). We thus conclude that the companion's extreme redness is not likely to be caused by the debris disk, nor is this level of extreme optical depth likely in a circumsecondary disk. Further effects of reddening due to the disk environment are explored in detail in Section~\ref{sec:environ}. The $A_{V}$ estimate of 0.03 from stellar SED modeling in Figure~\ref{fig:stellar_posteriors} also argues against a significant reddening effect due to interstellar extinction, leaving dust in the companion atmosphere as the most viable scenario. 

\subsection{Atmospheric reddening from dusty aerosols}
\label{sec:reddening}

The known L-dwarf population exhibits a wide range of atmospheric color in the NIR, spanning $\sim$1.5 magnitudes in $J-K_{s}$ \citep{faherty13}. In particular, young L-dwarfs associated with moving groups have redder NIR colors than field counterparts of the same spectral type, an effect attributed to lower surface gravities that retain clouds at high altitude \citep{marley12}. From a large survey of 420 ultracool dwarfs, \cite{kellogg17} found similar ($\sim$2\%) fractions of unusually red objects in both younger ($<$200~Myr) and older ($\geq$200~Myr) populations, an age delineation corresponding to the halting of gravitational contraction within ultracool evolutionary models \citep[e.g.,][]{burrows01, baraffe15} and roughly correspondent to the demarcation between very-low to intermediate gravity indicators from \cite{allersliu13}. While age and peculiarity of ultracool objects are difficult to precisely quantify, a significant population of reportedly older red L-dwarfs exists \citep[cf.][]{looper08}. Inclination angle has also been invoked as a potential explanation for discrepant red colors, as substellar objects viewed equator-on have redder infrared colors for the same spectral type than those viewed pole-on \citep{kirkpatrick10, vos17}.

Reasons for persistence of redder colors in moderate age, high-gravity objects remain uncertain. However, the presence of upper-atmosphere dusty aerosols of sub-micron sized grains has been put forth as a potential explanation. In this framework, small, cool dusty aerosols high in the atmosphere scatter (but do not emit) thermal flux. The sensitivity of scattering efficiency to wavelength for such particles is small compared to the wavelength of interest, thus producing a reddening. \citet{marocco14} and \citet{hiranaka16} examined such extinction effects in peculiar red L-dwarf atmospheres by assuming a population of sub-micron grains in the upper atmosphere. \citet{marocco14} adopted a de-reddening approach with wavelength-dependent corrections based on standard interstellar extinction laws \citep{cardelli89,fitzpatrick99}, and showed that applying dereddening using common dust compositions could make peculiar red objects appear similar to the field L-standard population. The HD~206893~B analysis by \citet{delorme17} identified the spectral type as L5--L7, concordant with its extremely red $Js-K1$ color, and they applied a similar method, generating extinction profiles with a range of A$_{v}$ (up to A$_{v}$=10) and particle sizes (from 0.05 to \SI{1}{\um}). However, corresponding objects in the same region of the NIR CMD exhibit significantly deeper water absorption features at \SI{1.4}{\um} and much bluer slopes than those evident in the spectrophotometry of HD~206893~B, regardless of youth. By comparing the de-reddened spectrum of HD~206893~B to standard objects using a Cushing G analysis, \citet{delorme17} determined the closest matches to be those of field or young L3.5 dwarfs, reproducing the slope of the SPHERE spectrophotometry.

\begin{figure*}[h]
    \centering
    \includegraphics[width=0.9\textwidth]{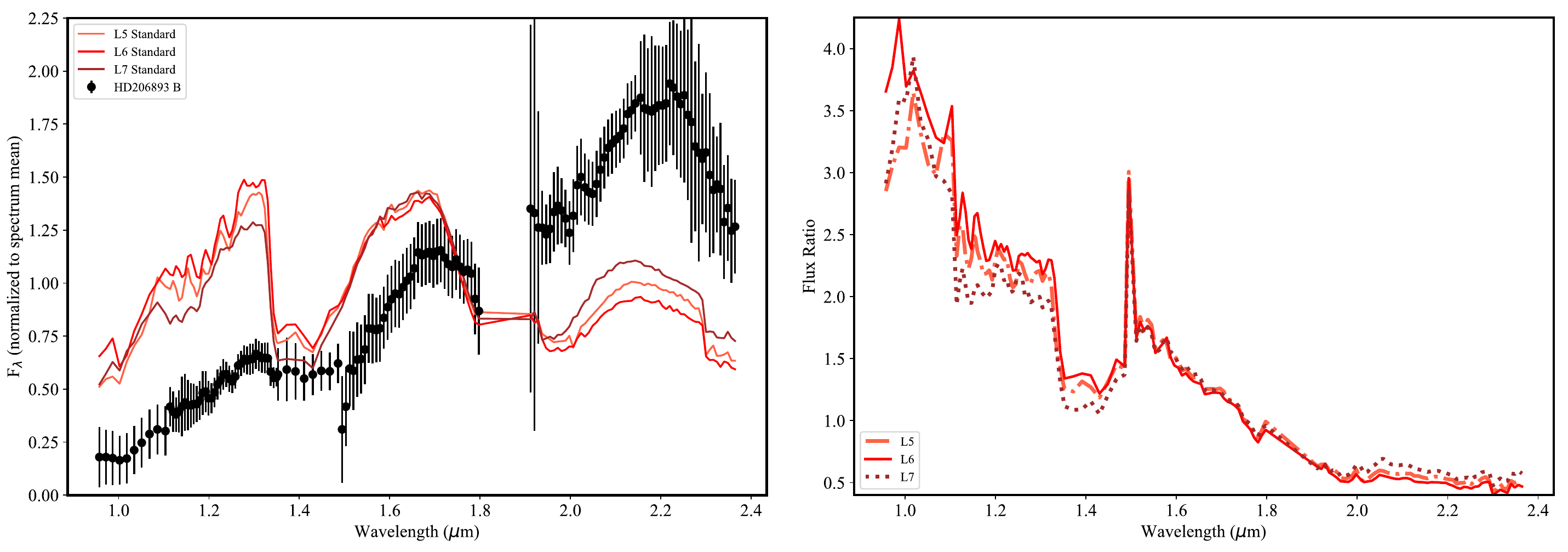}
    \caption{Comparison of the full SPHERE+GPI spectrum of HD~206893~B with three late field L-dwarfs (L5: 2MASS~J06244595-4521548; L6: 2MASSI~J1010148-040649; L7: 2MASSI J0825196+211552). The right panel plots the flux ratios of the standard spectra divided by that of HD~206893~B. The spike at $\sim$\SI{1.55}{\um} is an artifact corresponding to the gap between the SPHERE IFS band edge and the onset of GPI $H$-band coverage.}
    \label{fig:standardcomp}
\end{figure*}

In \citet{hiranaka16}, extinction profiles generated from Mie scattering models with various grain sizes, size distributions, and compositions were fit to extinction profile estimates derived by dividing spectra of unusually red L-dwarfs by those of typical field L-dwarfs. In this work, we apply the same methodology and compare the full NIR spectrum of HD~206893~B to known L-dwarfs in order to investigate potential small dust grain populations that may be responsible for the observed reddening.

Figure~\ref{fig:standardcomp} shows the full spectrum of HD~206893~B divided by L5, L6 and L7 spectral standards, selected as objects with median $J$-$Ks$ colors for their spectral classes from the field gravity template sample in \citet{cruz18}. Dividing the normalized spectrum of the standard by the normalized companion spectrum yields a visualization of the observed reddening, shown as the flux ratio of the two objects. The reddening profile is then fit by grain populations with various scattering properties, which may be used in turn to de-redden the companion spectrum. Initial treatments of dust in very low-mass stellar atmospheres recognized that the condensation temperatures of species such as enstatite (\ce{MgSiO3}), iron (Fe), forsterite (\ce{Mg2SiO4}), and corundum (\ce{Al2O3}) occurred in the photospheres of cool M-dwarfs \citep[e.g.,][]{tsuji96} and many additional grain compositions with cool condensation temperatures have since been incorporated into atmospheric models treating dust \citep[e.g., $\sim$30 various species of dust;][]{allard01}. As in \citet{hiranaka16}, we use Mie scattering models and the refractive indices of various species to determine extinction coefficients. We adopt a standard power-law distribution $n(a) = a^{-3.5}$ with grain sizes from 0.1-\SI{1}{\um} and use the Mie scattering code LX-MIE \citep{kitzmannheng18} to calculate extinction curves for particles of forsterite, corundum, and \ce{TiO2}. As all three species produced similar extinction profiles and grain parameters, we focus here on the results from the forsterite grain fitting. 

\begin{figure*}[h]
    \centering
    \includegraphics[width=0.9\textwidth]{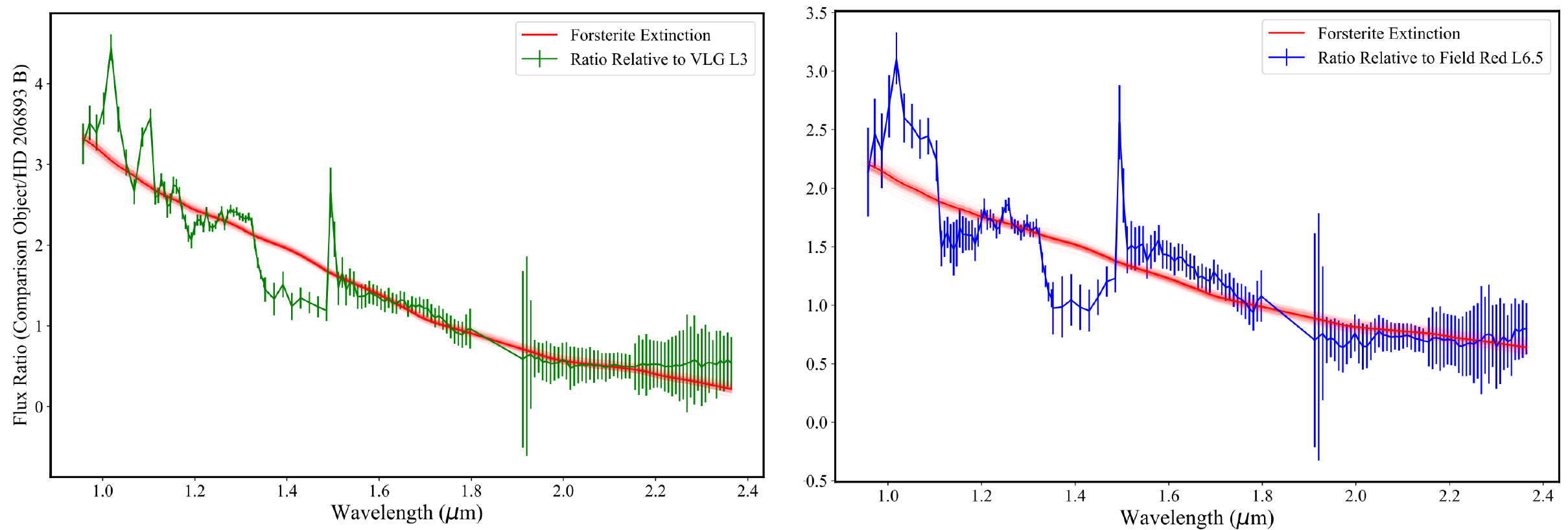}
    
    \caption{Best-fit forsterite grain aerosol models (red) with the corresponding parameters from the MCMC analysis shown in Figure~\ref{fig:mieposteriors}. The green curve represents the observed extinction for HD~206893~B when compared with a VLG L3 object, 2MASSW J2208136+292121 (left panel), and the blue curve for a red field L6.5pec, 2MASSW~J2244316+204343 (right panel).
    }
    \label{fig:extinctioncurve}
\end{figure*}

\begin{figure*}[h]
    \centering
    \includegraphics[width=1.0\textwidth]{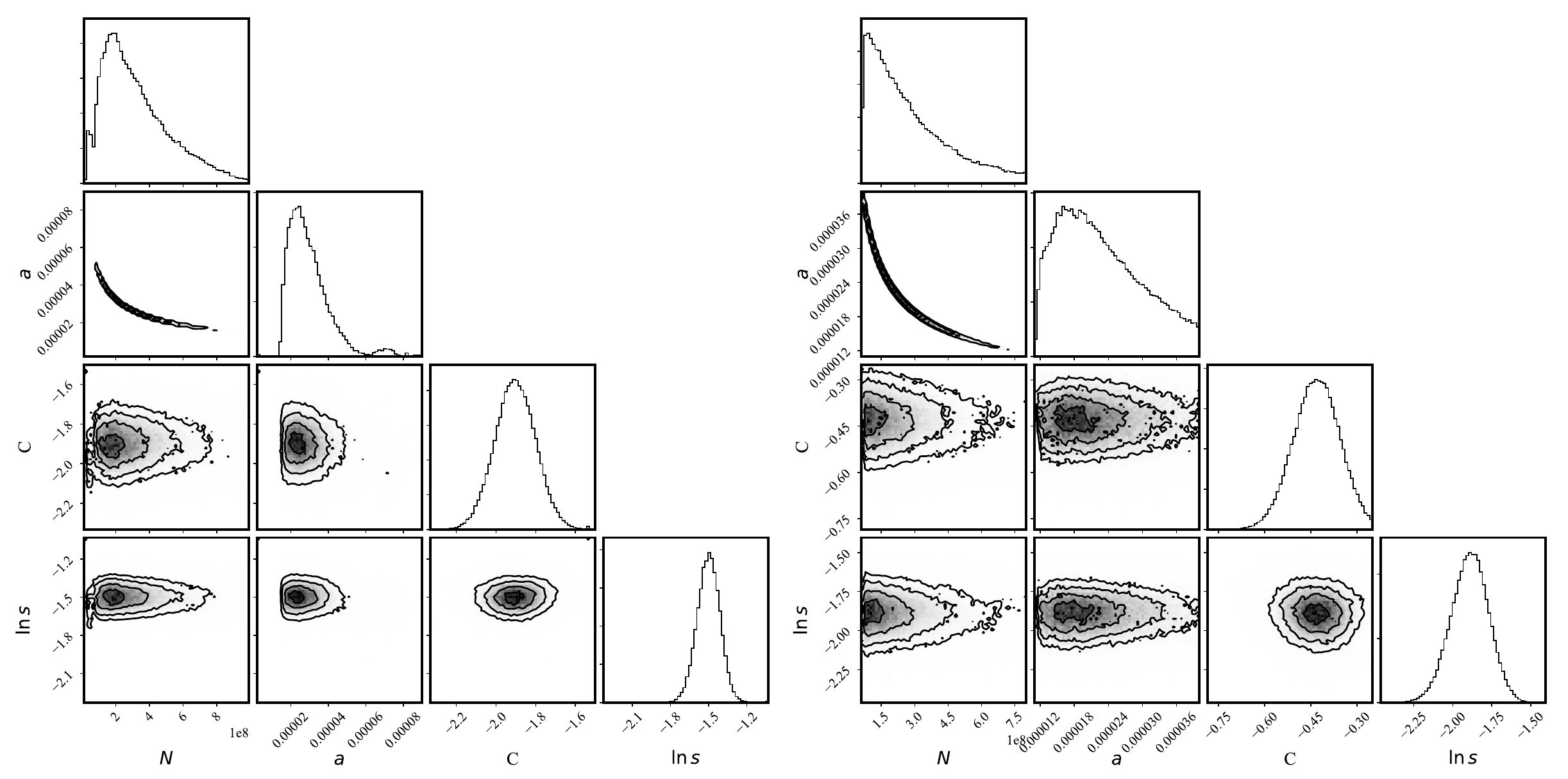}
    \caption{Posterior distributions for the best-fit extinction profiles in Figure~\ref{fig:extinctioncurve}, with the left panel corresponding to the fit to the assumed L3 extinction profile, and the right panel to the assumed L6.5pec extinction profile. Distributions shown include grain column density in $10^{8}$ particles per \si{\square\cm} ($N$), mean grain size in cm ($a$), vertical offset due to gray opacity ($C$) and an error tolerance parameter ($\ln(s)$). The covariance of the mean grain size and column density reflect the inverse proportionality of the parameters, with similar extinction properties for small grain sizes.}
    \label{fig:mieposteriors}
\end{figure*}

Given the previous \citet{delorme17} analysis demonstrating the closest spectral approximation of HD~206893~B to a reddened atmosphere of spectral type L3 (consistent with both field and younger AB Dor member L3 objects), we perform extinction fitting with an extinction curve derived from the flux ratio of HD~206893~B to a low-gravity L3$\gamma$ object \citep[2MASSW J2208136+292121;][]{kirkpatrick00}\footnote{We note that the L3$\gamma$ object J2208 was identified as a $\beta$ Pic candidate by \citet{gagne14}, while an updated parallax and CMD analysis by \citet{liu16} show that it is marginally more similar to the field CMD sequence than that of very low gravity objects, making its youth determination slightly uncertain; however, the latter authors conclude that it remains a promising candidate member of the young moving group.}, and an unusually red field object with NIR spectral type L6.5pec \citep[2MASSW~J2244316+204343;][]{dahn02,mclean03,looper08}. Fits to each of the extinction curves were performed using a Bayesian MCMC approach \citep[the \textit{emcee} package;][]{foremanmackey13} to estimate particle column density, mean grain radius, and opacity scaling. The best fit curves from the MCMC analysis are shown in Figure~\ref{fig:extinctioncurve}. For the earlier very low gravity L3 spectral type, the corresponding posterior distributions are shown in Figure~\ref{fig:mieposteriors}, with a best fit column density of \SI{2.8e8}{particles/cm^{2}}, mean particle radius of \SI{0.27}{\um}, and a constant vertical offset term of $C=-1.9$ which corresponds to a gray atmospheric opacity scaling term between the object and the field spectrum. For the peculiar L6.5, the best fit values have slightly lower column density (\SI{2.5e8}{particles/cm^{2}}), and smaller mean particle radius (\SI{0.20}{\um}), with a constant vertical offset term of $C=-0.43$.

\begin{figure}

    \centering

    \includegraphics[width=0.48\textwidth]{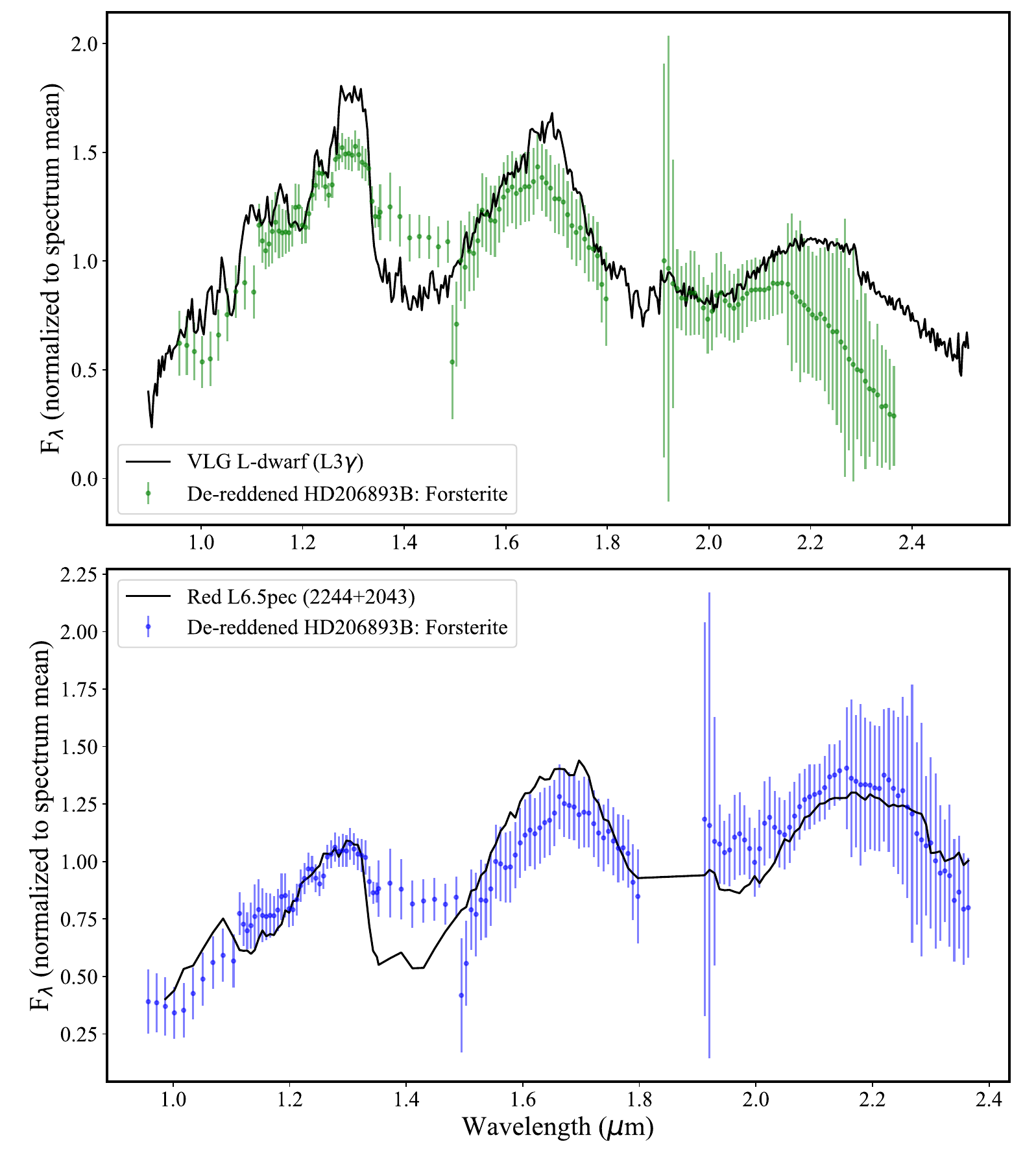}

    \caption{Dereddened spectra of HD~206893~B shown with spectra of a VLG L3 object, 2MASSW J2208136+292121 (top panel) and a red field L6.5pec, 2MASSW~J2244316+204343 (bottom panel). The de-reddening approach is capable of producing realistic $J,H$ and $K1$-band spectral features in either case, albeit with too shallow \ce{H2O} absorption, and is well-approximated by the peaky $H$-band morphology of the lower surface gravity L3 object. However, the $K2$ band deviates significantly from that of an earlier L-dwarf, and is more similar to that of a late red field object with additional extinction.}

    \label{fig:dereddenedL3}

\end{figure}

In Figure~\ref{fig:dereddenedL3}, we show the result of applying the best-fit extinction curve for forsterite to effectively ``de-redden'' the full spectrum of HD~206893~B and compare it both the low gravity L3 object and unusually red field L6. The de-reddened spectra correspond closely to the L6.5 over $J$-band and similarly to both the L3 and L6.5 over $H$ bands, albeit with a weaker water absorption feature at \SI{1.4}{\um}, and more closely replicates the peaky $H$-band shape of the lower-gravity L3 than the broader morphology of the L6.5. The shallowness of the water band is consistent with the presence of significant dust in the object atmosphere; for late-M and L-dwarfs, \citet{leggett01} noted that below $T_\textnormal{eff} < 2500$K, the presence of dust heats the atmosphere in the line-forming region, and the water features may become broader and shallower, depending upon the dust properties (e.g., metallicity). A detailed analysis of such feedback effects between the posited dust layer and the atmosphere are beyond the scope of this investigation.

However, while the blue slope of $K1$ is roughly similar between the L3 and the GPI spectra, the decline seen at the red end of $K2$ deviates sharply from the L3 object spectrum. In comparison, the dereddened spectrum assuming HD~206893~B is similar to an unusually red late field object (L6.5pec) provides a more similar match to the general morphology of $K$ band, with less significant departure from the template object at the longest wavelengths in $K2$. As a slight red slope departure appears to persist after the de-reddening application of a small particle extinction profile, if physical, its origin may be attributed to factors beyond a high-altitude aerosol layer.

\clearpage

\subsection{A possible atmospheric reddening scenario for HD~206893~B}
\label{sec:environ}
The high fractional infrared luminosity of HD~206893~A \citep[$L_{dust}/L_{*} = 2.3 \times 10^{-4}$;][]{moor06} and existence of a companion interior to its debris disk make its architecture similar to companion-disk systems such as HD~95086 \citep[$L_{dust}/L_{*} = 1.4 \times 10^{-3}$, companion at $\sim$56~au;][]{chen12,rameau16}, HR~2562 \citep[$L_{dust}/L_{*} = 1.1 \times 10^{-4}$, companion at 20~au;][]{moor06, konopacky16}, and most recently, HD~193571 \citep[$L_{disk}/L_{*} = 2.3 \times 10^{-5}$, companion at 11~au;][]{barcucci19}. Of these systems, all of which have notably red companions, HD~206893~B remains the reddest, despite the fact that the system does not have the highest fractional infrared luminosity. As described in Section~\ref{sec:reddening}, additional dust contributions (e.g., in the form of high-altitude aerosols) are necessary to reconcile the spectrum of HD~206893~B with that of field and low-gravity L-dwarfs. Here we examine the plausibility of accretion from the debris disk as a potential source of dusty material in the HD~206893~B atmosphere. 

As a simple approximation, the amount of dust required to redden the companion atmosphere can be compared to the debris disk properties derived from previous observations and modeling. We assume that any accreted dust remains solid, and is notfully vaporized during the accretion process\footnote{We note that the high temperature of an accretion shock similar to those seen in T Tauris could destroy dust, while the ablation and/or vaporization of incident meteoroids is often observed in the Solar System \citep[e.g., on the Earth and Mars;][respectively]{hunten80, hartwick19}, and are thought to seed high-altitude clouds. Also cf. the impact of comet Shoemaker-Levy~9, which produced debris aerosols in the Jovian atmosphere with particle size 0.13--\SI{0.3}{\um} \citep{west95}.}. From the derived extinction required to reconcile HD~206893~B's spectrum to that of a VLG L3 object in Section~\ref{sec:reddening}, the derived column density of $\sim$\SI{3e8}{cm^{-2}} can be used to estimate the excess dust contribution in an aerosol layer. Assuming that the object radius is $1~R_\textnormal{Jup}$, and that high-altitude aerosols consist of forsterite particles of size $\sim$\SI{0.3}{\um} and density \SI{3.27}{g/cm^3}, suggest the presence of $\sim$\num{9e-10}~$M_\textnormal{Moon}$ of dust within the atmosphere. Previous analyses of the HD~206893~B debris disk fit the SED with both single and two-component dust temperature models: the two-component dust temperature model included a hot (499~K) dust component at 0.8~au consisting of \num{1.1e-6}~$M_\textnormal{Moon}$ of material, and a colder (48~K) dust belt at 261~au of \num{5.6e-1}~$M_\textnormal{Moon}$ of material \citep{chen14}. The single-component models consisted of a blackbody of 54~K dust at 43~au, with an estimated mass of 0.030~$M_{\oplus}$ (2.4~$M_\textnormal{Moon}$) derived from \SI{850}{\um} JCMT/SCUBA observations \citep{holland17}. Adopting a conservative limit of \num{1.1e-6}~$M_\textnormal{Moon}$ of hot dust in the inner portion of the disk, we can roughly estimate relative timescales of dust accretion onto the substellar companion. 

Accretion rates of solids onto young planets have been estimated at $\sim$\num{e-6}$M_{\oplus}~$yr$^{-1}$ for gas-rich disks \citep[\num{8e-5}$M_\textnormal{Moon}$~yr$^{-1}$; e.g.,][]{alibert18}. Assuming this high solid accretion rate, the time required to collect enough material to redden HD~206893~B would be significantly less than one year. However, the HD~206893 debris disk is significantly more evolved, with a much lower dust surface density than younger gas-rich disks. We therefore examine two limiting cases for potential accretion rates: (1) Accretion of solids onto a giant planet based on orbital parameters, and consistent with estimates for younger protoplanetary disks \citep[\num{e-4}$M_\textnormal{disk}$/orbit;][]{paardekooper07}, and (2) Estimates of interplanetary dust flux at Jupiter derived from \textit{in situ} measurements from \textit{Pioneer 10} and the \textit{New Horizons} Student Dust Counter \citep{poppe16}.

In the high accretion rate (protoplanetary) scenario, we assume an orbital period of 32 years, as estimated in Section~\ref{sec:astrometry}, a dust mass of \num{1.1e-6}~$M_\textnormal{Moon}$ in the vicinity of the companion \citep[corresponding to the hot dust estimate from][]{chen14}, and a solid accretion rate of \num{e-4}$M_\textnormal{disk}$/orbit, which implies a mass accretion rate of \num{4e-12}$M_\textnormal{Moon}$~yr$^{-1}$. Using the derived column density for the companion, this suggests that only a very short period of time ($\sim$260 years) would be required to accrete sufficient dust from the environment and redden the atmosphere to the extent currently observed. In contrast, in the low accretion Solar System-like scenario, typical values of incident dust flux at Jupiter are on the order of \SI{e-13}{g.m^{-2}.s^{-1}} of 0.5 to \SI{100}{\um} material \citep{poppe16}, corresponding to a much lower accretion rate of \num{6e-16}$M_\textnormal{Moon}$~yr$^{-1}$ (for a Jovian-sized body), and requiring $\sim$1~Myr to accumulate the estimated dust content in HD~206893~B. 

The steady state reached by the atmosphere depends upon the lifetime of the accreted dust, which in turn depends upon the particle-size dependent fall speed and the strength of atmospheric eddy mixing. In lower gravity atmospheres, the fall speed is lower and the effect of a given strength of eddy mixing is stronger, all else being equal, favoring longer dust lifetimes. A complete analysis, accounting for the coupled problems of radiative heating of the atmosphere by the accreted dust and the dust sedimentation, will be considered in the future. 

Either of the two accretion scenarios is plausible considering the age of the system and potential replenishment of dust within the disk, and a significantly higher dust estimate at the $\sim$10~au separation of HD~206893~B than the conservative estimate adopted here would increase the interception of grains from the environment into the atmosphere. If the companion is indeed low-surface gravity, sustaining a small dust grain population at high altitude in the atmosphere may be possible for extended periods, particularly if debris disk dust production replenishes the aerosol layer. In comparison to the red colors of the young/low-gravity substellar population, which are already postulated to result from the presence of thick high-altitude clouds, the even redder color of HD~206893~B could potentially be attributed to an additional source of reddening from the dusty disk environment in which it resides.


\section{Conclusions} 
\label{sec:conclusion}

We present GPI spectroscopy of the substellar companion HD~206893~B, obtained at $J$, $H$, $K1$, and $K2$ bands. Consistent with the extraordinary red nature at $H-L'$ initially noted by \citet{milli17}, the broader spectral coverage of GPI further supports its exceptionally red nature. The overlapping wavelength regimes between the GPI and SPHERE observations shows excellent agreement with the SPHERE $YJH$ observations presented by \citet{delorme17}. The addition of full $H$ and $K1/K2$ spectroscopy made possible with GPI suggests that the companion may have low surface gravity and that its spectrum appears morphologically most similar to that of the young low-gravity late L-dwarf population. 

The GPI photometry of HD~206893~B consistently demonstrates its extraordinary position in color-magnitude space, with $H-Ks=1.68\pm0.08$, significantly redder than the previously reddest substellar object, 2MASS~J22362452+4751425~b \citep[2M2236b, $H-K=1.26\pm0.18$,][]{bowler17}. From comparison of the GPI spectra to brown dwarf spectral libraries, we find that the closest matching spectrum is that of a late-type, possibly low-gravity L-dwarf, enabling comparison with other extremely red/peculiar objects like 2M2236b (``late-L pec''), 2M1207b \citep[M8.5-L4;][]{patience10}, and PSO~J318.5-22 \citep[L7;][]{liu13}. 

We apply a de-reddening approach akin to that of \citet{hiranaka16} to determine whether a small grain, sub-micron aerosol layer could reconcile the observed spectrum with that of field and low-gravity L-dwarfs. We find that for reasonable column densities and grain properties, both a low gravity L3 and ``peculiar'' red field L6.5 provide good matches to HD~206893~B, with the overall spectrum matching that of the later-type object but the spectral shape of features in, e.g., $H$-band more closely approximated by that of the lower-gravity object.

The emergent spectrum of HD~206893~B proves challenging to fit when conducting comparisons with a suite of various atmospheric model grids, owing to its enhanced luminosity at $K$-band relative to shorter wavelengths and a slightly unusual $K2$ morphology. Conducting a model grid fit to each of the GPI bands separately provides different companion parameters as compared to fitting the full $Y$ to $L$ spectral coverage of SPHERE, GPI, and NaCo. Each of the four model grids provided more internally consistent effective temperatures for the individual bands (ranging from 1400~K to 1800~K, with an average across all bands and grids of $\sim$1540~K), albeit over the full range of log($g$)=3.5--5.0, making the surface gravity of the object ambiguous.

We provide uncertainties on the stellar age ranging from 40-600~Myr, adopting 250~Myr for completeness analyses, but highlight that the high infrared luminosity of the disk and the non-negligible ($61-63\%$) likelihood of membership in the Argus association point to a younger age for the system. As the mass of the companion depends on the assumed age, the companion mass ranges from 12-40 M$_\textnormal{Jup}$ for ages in the $50-500$~Myr range of COND models. The peaky morphology of the $H$ band spectra, good fit of the de-reddened spectrum to that of a late-type L-dwarf, morphological similarity of the spectrum to known young moving group late L-dwarfs, and the potential lower dynamical mass of 10~$M_\textnormal{Jup}$ estimated by \citet{grandjean19}, all also point toward a self-consistent scenario for HD~206893~B being significantly younger and lower-mass than initial age estimates and its luminosity have implied.

We have combined the five epochs of VLT astrometry presented in \citet{milli17}, \citet{delorme17}, and \citet{grandjean19} with four new multiband GPI epochs, spanning four years of orbital coverage. We estimate the orbital period to be $29.1^{+8.1}_{-6.7}$~years, with a probable semi-major axis of $10.4^{+1.8}_{-1.7}$~au and orbital inclination of $145.6^{+13.8}_{-6.6}$\,deg. From previous estimates of the debris disk inclination and inner gap radius of $\sim$50~au \citep{milli17}, these data are consistent with the companion being well-within the inner edge of the dust emission \citep[as previously suggested by][]{delorme17}. The potential significant gap between the $\sim$10~au best-fit semi-major axis and the 50~au inner disk radius has motivated the search for additional companions within the gap. However, no additional companions are detected in our images. The GPI data are 25--75\% complete to 5~M$_\textnormal{Jup}$ at orbital separations of 20-30~au, respectively (assuming an older system age of 250~Myr). If the system is significantly younger, this may suggest that any additional companion responsible for carving the disk edge would be sub-Jovian. Comparing our MCMC fits to the visual orbit with the initial disk geometry estimates from \citet{milli17} favors less misaligned and more co-planar orbital configurations; however, the measurement precision on disk and orbital elements cannot currently exclude moderately misaligned ($i_{m}\sim20^{\circ}$) configurations. As recent work has shown that $i_{m}$ correlates with the orbital period of companions in circumbinary systems \citep{czekala19}, with tighter binaries exhibiting greater coplanarity, this system provides a useful laboratory for disk-companion dynamics. \citet{milli17} caveat that the marginally-resolved nature of the debris disk from 70~\si{\um} \emph{Herschel} observations do not tightly constrain the disk geometry, necessitating future disk observations at higher spatial resolution and longer-term orbital monitoring to explore the dynamical interplay between the companion and disk environment. 

As the reddest substellar companion identified to date, HD~206893~B adds to the growing population of remarkably red planetary mass objects and free-floating objects. With its exceptional color and close orbital separation, HD~206893~B is an important comparison for similarly red free-floating objects (e.g., PSO~318), red, wide substellar objects like 2M2236b, and planetary-mass companions posited to have high photospheric dust content like HD~95086~b \citep{derosa16}. The redness of the companion and its extremely dusty nature also make it well-suited to exploring the effects of clouds, metallicity, and disequilibrium chemistry at cool temperatures in atmospheric models. Interpreting these observations, and future observations extended further into the infrared (e.g., with the \textit{James Webb Space Telescope}), will likely be subject to how dusty clouds, collisionally-induced absorption, and atmospheric chemistry are treated in models \citep{baudino17}, and unusual systems such as HD~206893~B will provide important laboratories for these processes. \textbf{Furthermore, if the system is indeed young, this companion would present an exceptional addition to the growing population of directly-imaged giant exoplanets.} If the system is $\sim$250~Myr, HD~206893~B represents only the second detection of a brown dwarf orbiting within the inner gap of its host debris disk \citep[after HR~2562;][]{konopacky16}, and is otherwise one of only six substellar companion systems with both resolved disks and companions. This makes it a key testing ground to explore dynamical interaction between disks, companions, and their host stars. The dustiness of the system also merits a search for polarized signal from the companion \citep[e.g., as observed in companion systems like CT Cha and HD~142527;][]{ginski18,rodigas14}. With multi-wavelength spectral coverage, detailed modeling, and continued dynamical monitoring, constraining the physical properties of this system will provide critical context for our understanding of the atmospheric and evolutionary histories of substellar objects.

\acknowledgments

We thank the anonymous referee for a exceptionally thoughtful and comprehensive report which greatly benefited this manuscript. K.W.D. thanks Alan Jackson for valuable insight on debris disk dust properties, and Brian Svoboda, Sarah Betti, Mike Petersen, Daniella Bardalez-Gagliuffi, Sarah Logsdon, and Emily Martin for helpful discussions and feedback. This work is based on observations obtained at the Gemini Observatory, which is operated by the Association of Universities for Research in Astronomy, Inc., under a cooperative agreement with the NSF on behalf of the Gemini partnership: the National Science Foundation (United States), the National Research Council (Canada), CONICYT (Chile), Ministerio de Ciencia, Tecnologia e Innovacion Productiva (Argentina), and Ministerio da Ciencia, Tecnologia e Inovacao (Brazil). This research has made use of the SIMBAD and VizieR databases, operated at CDS, Strasbourg, France. This research used resources of the National Energy Research Scientific Computing Center, a DOE Office of Science User Facility supported by the Office of Science of the U.S. Department of Energy under Contract No. DE-AC02-05CH11231. This work used the Extreme Science and Engineering Discovery Environment (XSEDE), which is supported by National Science Foundation grant number ACI-1548562. This work has made use of data from the European Space Agency (ESA) mission {\it Gaia} (\url{https://www.cosmos.esa.int/gaia}), processed by
the {\it Gaia} Data Processing and Analysis Consortium (DPAC, \url{https://www.cosmos.esa.int/web/gaia/dpac/consortium}). Funding for the DPAC has been provided by national institutions, in particular the institutions participating in the {\it Gaia} Multilateral Agreement. This work was supported by NSF grants AST-1411868 (K.W.D., A.R., J.P., B.M., E.L.N., and K.B.F.), AST-141378 (G.D.), and AST-1518332 (R.D.R., J.J.W., T.M.E., J.R.G., P.G.K.). Supported by NASA grants NNX14AJ80G (E.L.N., S.C.B., B.M., F.M., and M.P.), NNX15AC89G, NNH15AZ59I (D.S.), and NNX15AD95G (B.M., J.E.W., T.M.E., R.J.D.R., G.D., J.R.G., P.G.K.), NN15AB52l (D.S.), and NNX16AD44G (K.M.M). K.W.D. was supported by an NRAO Student Observing Support Award (SOSPA3-007). J.R., R.D. and D.L. acknowledge support from the Fonds de Recherche du Quebec. J.R.M.'s work was performed in part under contract with Caltech/JPL funded by NASA through the Sagan Fellowship Program executed by the NASA Exoplanet Science Institute. Support for M.M.B.'s work was provided by NASA through Hubble Fellowship grant \#51378.01-A awarded by the Space Telescope Science Institute, which is operated by the Association of Universities for Research in Astronomy, Inc., for NASA, under con- tract NAS5-26555. J.J.W. is supported by the Heising-Simons Foundation 51 Pegasi b postdoctoral fellowship. Portions of this work were performed under the auspices of the U.S. Department of Energy by Lawrence Livermore National Laboratory under Contract DE-AC52-07NA27344. This work benefited from NASA's Nexus for Exoplanet System Science (NExSS) research coordination network sponsored by NASA's Science Mission Directorate.

\appendix

\section{Pipeline post-processing and covariance estimation}
\label{sec:appendix_pyklip_covar}
\subsection{pyKLIP Reduction}
\begin{figure}[h]
    \centering
    \includegraphics[width=0.75\textwidth]{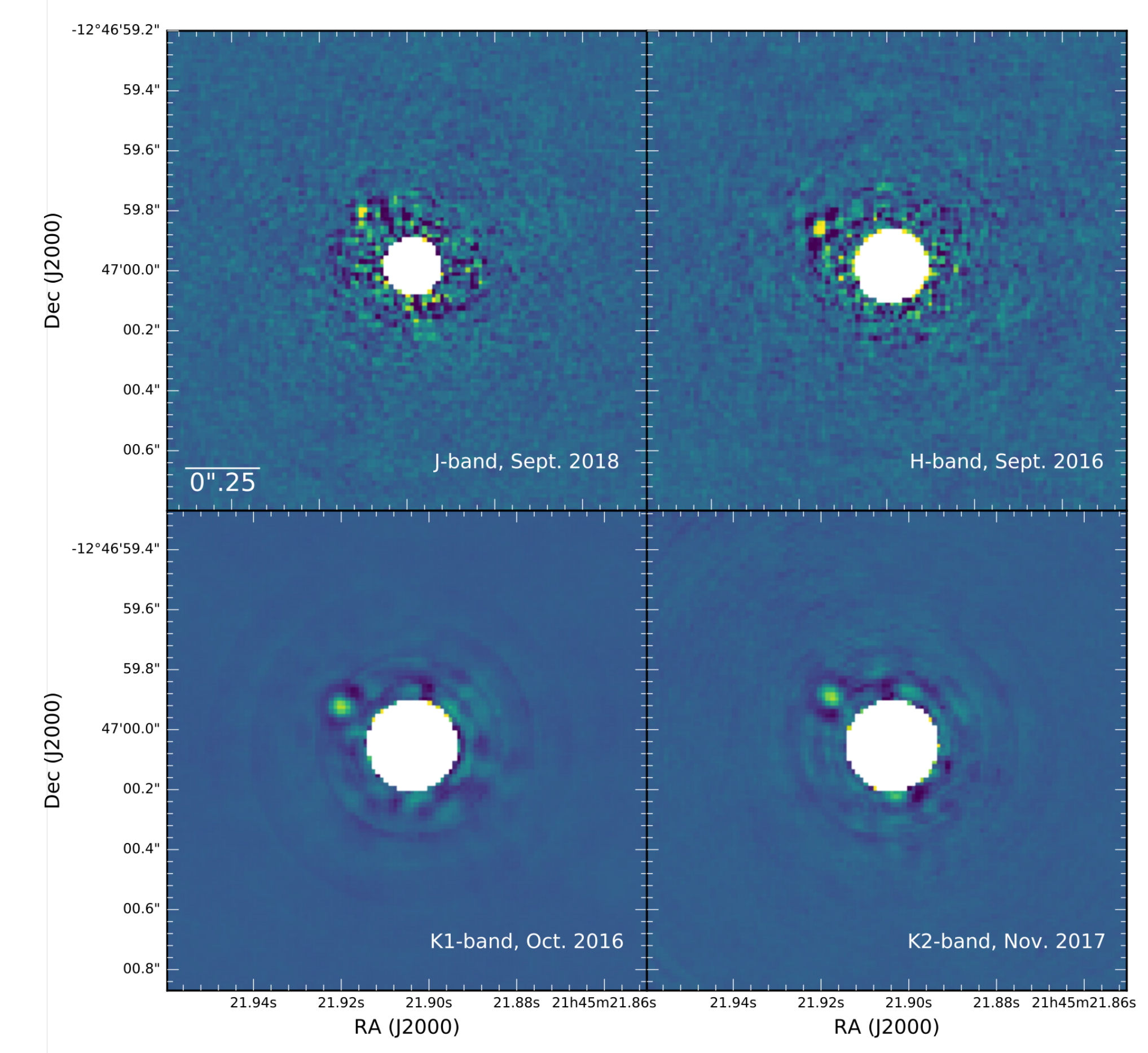}
    \caption{Auto-reduced and pipeline-processed PSF-subtracted images of HD~206893~B in $J$-band (upper left), $H$-band (upper right), $K1$-band (lower left), and $K2$-band (lower right). Images shown are from the pyKLIP reductions, oriented north up and east left, with parameters of 9 annuli, 4 subsections, 1 pixel movement criterion, and using the first 10 KL modes. Each individual image is scaled separately to minimize the contributions of residual speckle noise. Noticeable change in the companion position angle can be seen between the earliest 2016 epoch (upper right) and the latest 2018 epoch (upper left).}
    \label{fig:gallery}
\end{figure}

The images from the GPI auto-reduced pyKLIP pipeline \citep{wang18} are shown in Figure~\ref{fig:gallery}, with reduction parameters of nine annuli, four subsections per annulus, and exclusion criterion corresponding to movement of one pixel, using the first 10 KL modes to generate the reference PSF. The images show the close separation of the companion to the coronagraphic mask. 

\clearpage

\subsection{$K2$ Spectral Extraction}
\begin{figure}[h]
    \centering
    \includegraphics[width=0.75\textwidth]{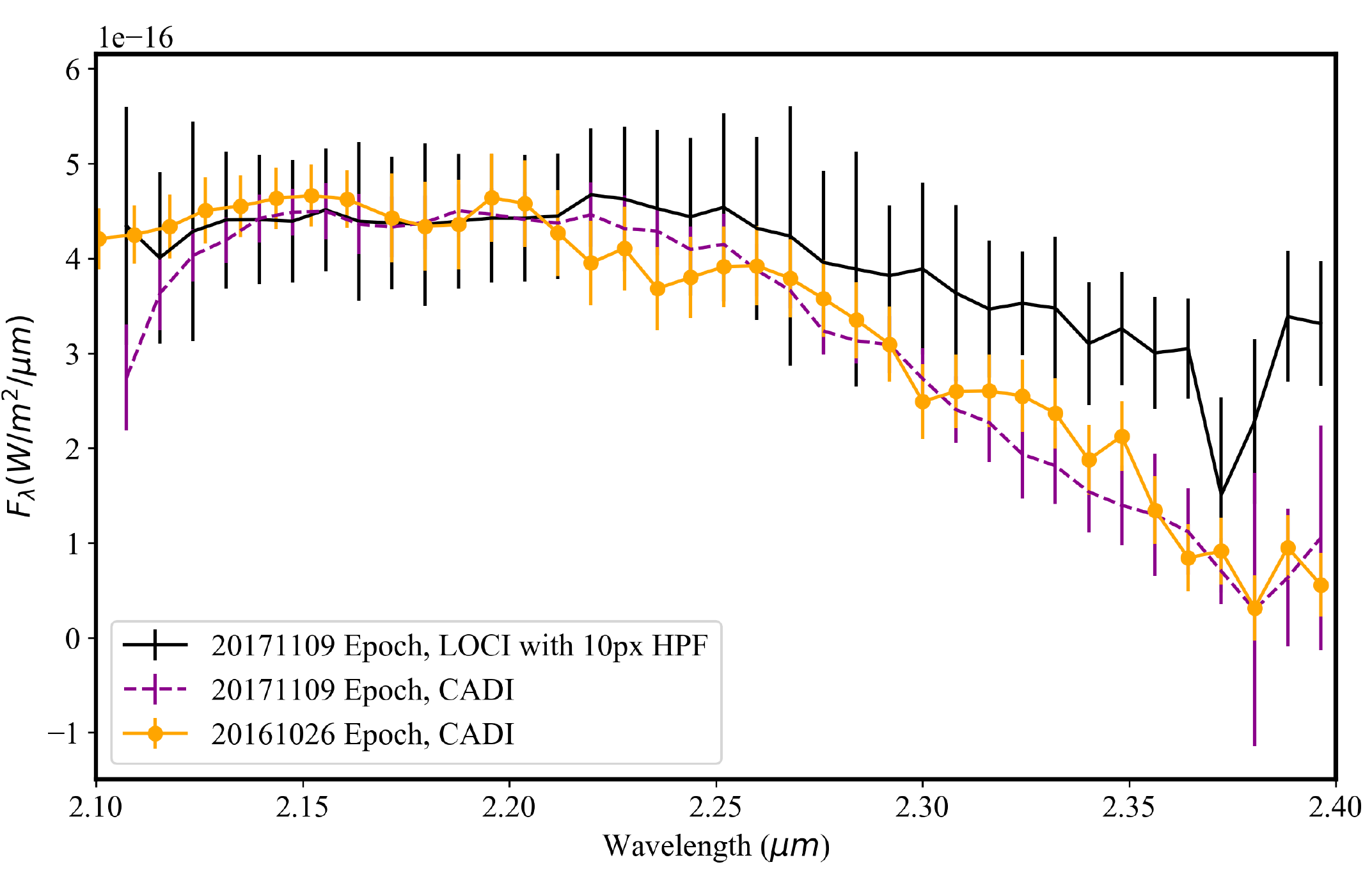}
    \caption{Comparisons of post-processing approaches for the GPI $K2$-band data, demonstrating the recovery of flux at longer wavelengths. The 2017-11-09 LOCI extraction (black line) with a broader high pass filter is used in this study.}
    \label{fig:k2comparison}
\end{figure}

As noted in Section~\ref{sec:results_spec}, the red end of the GPI $K2$ spectrum shows a steep spectral slope beyond 2.2$\mu$m. This slope was observed in both epochs of $K2$ data taken one year apart, and appears to be affected predominantly by the very low SNR in the GPI satellite spots at the red end of the band, which are critical to calibrate the measured flux in individual frames. The spectral slope is also impacted by the excessive brightness of the companion in these frames, the flux of which was significantly altered by the narrow high pass filter. Therefore, the shape of the extracted spectrum varies significantly depending upon decisions made in post-processing and PSF subtraction. We show the results of the 2016 CADI extraction in Figure~\ref{fig:k2comparison}, keeping in mind the strong bias of this spectrum due to very low SNR of the satellite spots, compared with the 2017 CADI and LOCI extractions with a high pass filter of 4 equivalent-pixels. Reducing the 2016 dataset with both LOCI and a broader high pass filter resulted in the same spectrum as with cADI and a narrow filter (4 pixels). The 2017 epoch of data has higher SNR owing to significantly more field rotation and longer integration, and because LOCI performed with a moderately large high-pass filter (10 pixels) recovers more flux from the extended wings of the PSF in the redder channels, we use the LOCI extraction throughout the analysis in this study.

\subsection{Covariance Estimation}
\begin{figure}[h]
    \centering
    \includegraphics[width=0.6\textwidth]{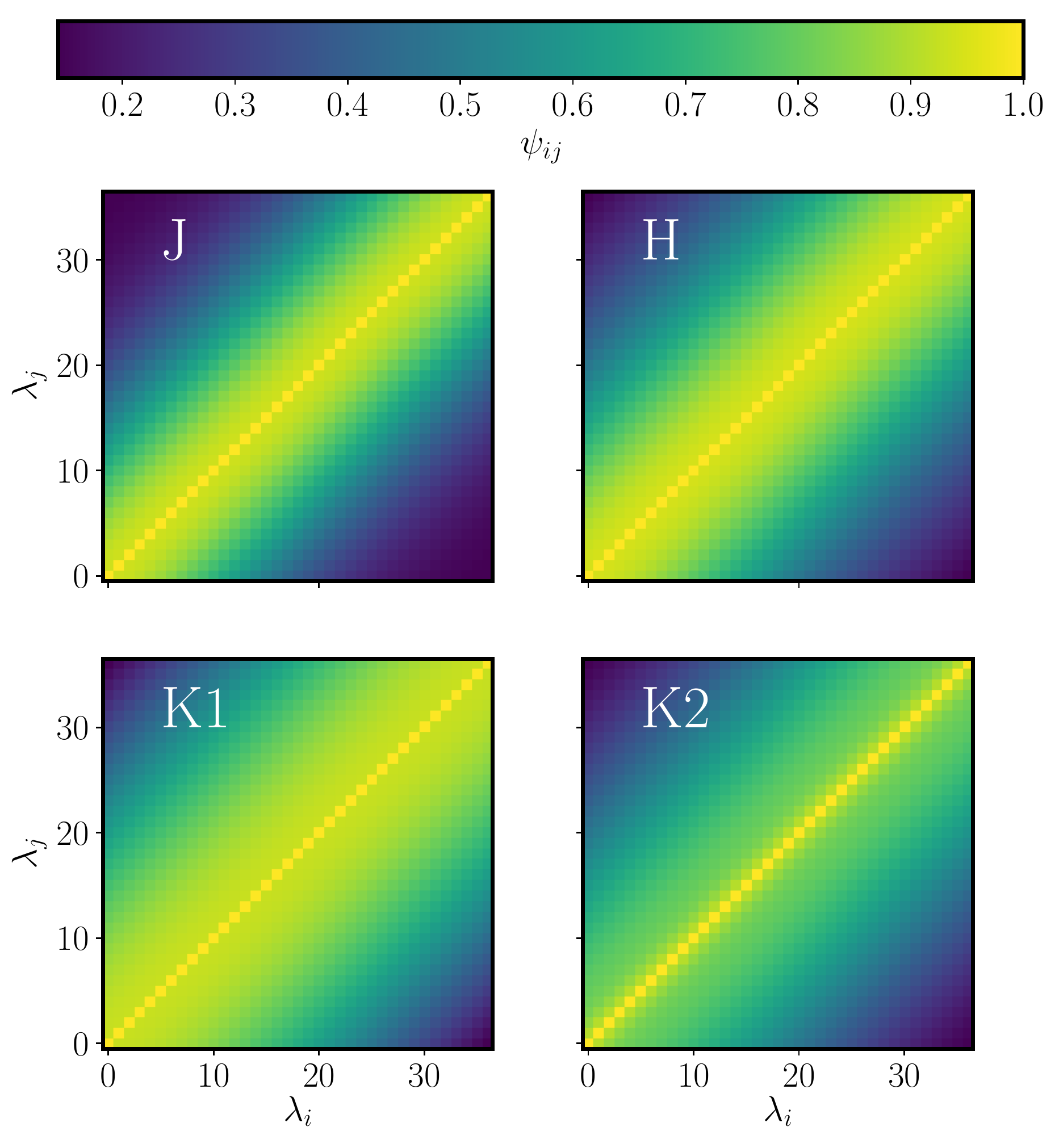}
    \caption{Spectral correlation matrices calculated from the LOCI and cADI PSF-subtracted datacubes, with the correlation between pixel values at different wavelength slices shown as intensity. The greatest spectral covariance can be seen at $H$ and $K1$ bands, with broad off-diagonal terms corresponding to high correlation between adjacent wavelength channels.}
    \label{fig:covariance}
\end{figure}

Figure~\ref{fig:covariance} shows the spectral covariance matrices, derived from measuring inter-pixel correlation within the final PSF-subtracted LOCI and cADI images. Each covariance matrix was calculated independently for each GPI band by use of a parallel-tempered MCMC to fit the 18-parameters of spectral correlation in the IFS datacubes as outlined in \citet{grecobrandt16} and applied in \citet{derosa16}. Each MCMC was run with 128 walkers at 16 different temperatures and run for 5000 steps saving every tenth step, burning-in the first 1000 steps in the chains. Plotted are the spectral correlations as a function of wavelength channel for the final datacube in each GPI band, where the off-diagonal elements correspond to correlated noise terms. For each spectrum, the high and low-frequency noise components, corresponding to read/background noise and speckle noise, respectively, are extracted and introduced into the error budget of the final spectra separately. Table~\ref{tab:corrampslengths} provides the fitted amplitudes of the correlated noise terms with respect to angular separation (A$_{\rho}$) and with respect to wavelength due to interpolation or crosstalk (A$_{\lambda}$). Detailed results from applying the spectral covariance to model fitting of the spectra are provided in Sections~\ref{sec:modeling} and~\ref{sec:appendix_models}.

\begin{table}[]
\begin{tabular}{cccc}
\hline
\hline
Band                                                 & $\rho$ (mas) & A$_{\rho}$ & A$_{\lambda}$ \\
\hline
J ($\sigma_{\rho}$=0.51,   $\sigma_{\lambda}$=4.05)  & 200              & 0.70       & 0.25          \\
                                                     & 225              & 0.80       & 0.14          \\
                                                     & 250              & 0.57       & 0.33          \\
                                                     & 265              & 0.51       & 0.38          \\
                                                     & 280              & 0.75       & 0.14          \\
                                                     & 300              & 0.89       & 0.02          \\
                                                     & 325              & 0.89       & 0.01          \\
                                                     & 350              & 0.86       & 0.01          \\
                                                     
\hline                                                
H ($\sigma_{\rho}$=1.14,   $\sigma_{\lambda}$=0.07)  & 200              & 0.59       & 0.37          \\
                                                     & 225              & 0.60       & 0.34          \\
                                                     & 250              & 0.56       & 0.39          \\
                                                     & 265              & 0.70       & 0.23          \\
                                                     & 280              & 0.68       & 0.25          \\
                                                     & 300              & 0.55       & 0.37          \\
                                                     & 325              & 0.37       & 0.51          \\
                                                     & 350              & 0.04       & 0.81          \\
\hline                                                     
K1 ($\sigma_{\rho}$=0.37,   $\sigma_{\lambda}$=0.13) & 200              & 0.02       & 0.93          \\
                                                     & 225              & 0.73       & 0.18          \\
                                                     & 250              & 0.73       & 0.14          \\
                                                     & 265              & 0.63       & 0.22          \\
                                                     & 280              & 0.66       & 0.16          \\
                                                     & 300              & 0.74       & 0.01          \\
                                                     & 325              & 0.64       & 0.05          \\
                                                     & 350              & 0.63       & 0.01          \\
\hline                                                     
K2 ($\sigma_{\rho}$=0.31,   $\sigma_{\lambda}$=0.01) & 200              & 0.83       & 0.10           \\
                                                     & 225              & 0.82       & 0.10           \\
                                                     & 250              & 0.79       & 0.10           \\
                                                     & 265              & 0.78       & 0.10           \\
                                                     & 280              & 0.77       & 0.08          \\
                                                     & 300              & 0.65       & 0.17          \\
                                                     & 325              & 0.54       & 0.22          \\
                                                     & 350              & 0.51       & 0.26\\  
\hline                                                     
\end{tabular}
\caption{Measured amplitudes of the correlated noise terms for each GPI datacube (A$_{\rho}$ and A$_{\lambda}$), with the relevant correlation lengths measured for each band ($\sigma_{\rho}$, $\sigma_{\lambda}$). The separations within the datacube are selected to span the range of companion separations at various epochs of observation. \label{tab:corrampslengths}} 
\end{table}

\clearpage

\section{Detailed Results for Model Atmospheric Grids}
\label{sec:appendix_models}

\subsection{Modeling Results With and Without Covariances}
Section~\ref{sec:modeling} describes the modeling results implementing spectral covariance as summarized in Figure~\ref{fig:covariance} and Table~\ref{tab:corrampslengths}, and here we provide additional details on each of the model grids explored in the fitting of the HD~206893~B spectrum and the results from a standard $\chi^{2}$ minimization. Table~\ref{tab:modelresults} summarizes the properties of the best fit models to the full spectrum and all four bands, demonstrating the wide range of derived physical properties dependent upon the wavelength range, choice of models, and whether or not covariances were implemented. As noted in Section~\ref{sec:modeling}, the best fit models accounting for spectral covariances do not always pass through the spectral datapoints in cases where the spectra are highly correlated; however, the resulting derived parameters across all four bands using covariances agree more closely than the wider range of temperatures, surface gravities, and radii estimated from a standard $\chi^{2}$ minimization.

\begin{table}[]
\footnotesize
\centering
\begin{tabular}{l|c|cccc|cccc}
\hline
\hline

\multicolumn{2}{c}{ }                            & \multicolumn{4}{c}{Standard $\chi^{2}$ Fitting}   & \multicolumn{4}{c}{Fitting Incorporating Covariance} \\
\hline        

Model Grid    & Range & T$_\textrm{Eff}$ (K) & log(\textit{g}) & Radius (R$_\textnormal{Jup}$) & $\chi^{2}$ & T$_\textrm{Eff}$ (K)   & log(\textit{g})   & Radius (R$_\textnormal{Jup}$)   & $\chi^{2}$   \\
\hline
Sonora        & Full         & 1100     & 5.0    & 1.68    & 4.4        & 1200       & 5.0   & 1.36    & 3.5          \\
              & J            & 1200     & 5.0    & 1.19    & 0.3        & 1400       & 5.0   & 0.82    & 1.4          \\
              & H            & 1300     & 3.7    & 1.06    & 0.9        & 1400       & 3.7   & 1.13    & 1.4          \\
              & K1           & 1400     & 3.7    & 1.28    & 0.2        & 1400       & 3.7   & 1.28    & 2.7          \\
              & K2           & 1400     & 3.7    & 1.28    & 0.2        & 1500       & 3.7   & 1.18    & 0.7          \\
\hline      
CloudyAE-60   & Full         & 1200     & 3.5    & 1.67    & 4.4        & 1700       & 3.5   & 1.55    & 8.6          \\
              & J            & 1700     & 3.5    & 1.31    & 1.8        & 1700       & 3.5   & 1.25    & 5.2          \\
              & H            & 1400     & 3.5    & 2.56    & 2.0        & 1700       & 3.5   & 1.77    & 2.4          \\
              & K1           & 1700     & 3.5    & 2.68    & 0.5        & 1700       & 3.5   & 2.82    & 7.3          \\
              & K2           & 1700     & 3.5    & 2.77    & 0.2        & 1700       & 3.5   & 2.80    & 1.1          \\
\hline              
DRIFT-PHOENIX & Full         & 1400     & 4.5    & 1.68    & 0.8        & 1400       & 4.5   & 1.36    & 1.2          \\
              & J            & 1400     & 3.0    & 1.23    & 0.1        & 1500       & 4.5   & 1.33    & 0.9          \\
              & H            & 1100$^{*}$     & 5.0    & 3.01    & 0.2        & 1400       & 3.0   & 1.34    & 0.9            \\
              & K1           & 1800$^{*}$     & 3.0    & 0.81    & 0.2        & 1800$^{*}$       & 3.5   & 0.80    & 2.5            \\
              & K2    & 2000     & 5.0    & 0.71    & 0.2        & 1300$^{*}$       & 5.0   & 1.63    & 0.6          \\
\hline              
BT-Settl      & Full         & 1600     & 3.5    & 0.94    & 1.6        & 1600       & 3.5   & 0.96    & 1.2          \\
              & J            & 1600     & 3.5    & 0.91    & 0.1        & 1600       & 3.5   & 0.88    & 0.9          \\
              & H            & 1600     & 3.5    & 0.87    & 0.3        & 1500       & 3.0   & 1.12    & 1.1          \\
              & K1           & 1800     & 4.5    & 0.76    & 0.1        & 1700       & 4.0   & 0.80    & 2.2          \\
              & K2           & 1700     & 4.5    & 0.83    & 0.2        & 1500       & 3.5   & 1.28    & 0.7  \\

\hline              
\end{tabular}
\caption{Best-fit model properties for each of the four suites of models, both using a standard $\chi^{2}$ minimization (visualized in Figure~\ref{fig:all_best_models}) and when incorporating covariance matrices. Asterisks ($^{*}$) indicate best-fit models from DRIFT-PHOENIX with super-solar metallicity ($[M/H] = +0.3$). \label{tab:modelresults}} 
\end{table}

\subsection{Model Descriptions}
\textbf{\textit{Sonora Models}:}
We conducted model comparisons with a subset of cloudy, solar-metallicity models from the upcoming Sonora 2020 model grid (Marley et al. 2020, in prep.). The Sonora models are applicable to brown dwarfs and giant exoplanets and incorporate revised solar abundances \citep{lodders10}, rainout chemistry, updated opacities for \ce{H2}, \ce{CH4} and alkali species, equilibrium and disequilibrium chemistry, and span both cloudless and cloudy models with temperatures of 200--2400~K and metallicities of $-0.5\leq [M/H] \leq2$. The subset of solar metallicity cloudy models tested here span $T_{eff}$=1100--1600~K and log($g$)=3.75--5.0. The cloudy Sonora models use the cloud model of \citet{ackermanmarley01} which is parametrized by the grain sedimentation efficiency, $f_{sed}$. This parameter sets the balance between grain sedimentation and upward motion from turbulent mixing. Small values of $f_{sed}$ correspond to slow sedimentation, smaller grains, and vertically extended clouds. The value of $f_{sed}$ was set equal to 1 (moderate particle settling efficiency) in the tested model subset, corresponding to a vertically extended thick cloud layer. Fitting over the full wavelength range favors a very low temperature 1200~K object with high surface gravity, and a physically-plausible radius of $\sim$1.36R$_\textrm{Jup}$. The results from fitting individual bands favor slightly higher temperatures and generally lower surface gravities, suggesting that the low temperature over the full wavelength range is driven by both spectral correlation and increasing redness in the relative band-to-band flux. These models more closely approximate the emergent flux across shorter wavelengths and the shallow water absorption features between $J$ and $H$, but underestimate the flux at the longest wavelengths.

\textbf{\textit{DRIFT-PHOENIX Models}:}
We also compared the spectra of HD~206893~B with the DRIFT-PHOENIX Atmosphere Models, which incorporate small grains in their mineral cloud physics prescription. The DRIFT-PHOENIX models are built upon the PHOENIX radiative transfer atmosphere models \citep{hauschildt92}, but also incorporate microparticle physics through the DRIFT code \citep{dehn07, helling08, witte09}, which includes the motion of particles within forming clouds and treats a variety of particle sizes and compositions. The DRIFT-PHOENIX models tested range from $T_{eff}$=1000--3000~K, log($g$)=3.0--6.0, and metallicities of solar $[M/H]$=0 or +0.3 (super-solar). For model comparisons with DRIFT-PHOENIX, solar metallicity models generally produced better fits than super-solar, concordant with the metallicity estimate for HD~206893~A from \citet{delorme17}; only two individual GPI bands were best-fit with $[M/H]$=+0.3 in either the standard $\chi^{2}$ or covariance fitting. The best model fit to the full spectrum favors a moderate temperature (1400~K) and surface gravity (log($g$) = 4.5), and plausible object radius (1.36R$_\textrm{Jup}$). The DRIFT-PHOENIX best-fit model most closely reproduces the general rise in flux from the bluest to reddest wavelengths in this study but underestimates the flux over $K$, in addition to shallower water absorption and overestimation of the $L'$ photometry.

\textbf{\textit{Cloudy-AE60 (Madhusudhan et al. 2011) Models}:}
Given the red nature of the companion, we fit its spectra with the publicly-available cloudy atmosphere models generated by \cite{madhu11} for the HR~8799 planets, which are also anomalously red relative to the field population. The full model grid includes both cloudy and cloud-free atmospheres with a range of forsterite cloud thicknesses and optical depths. The \cite{madhu11} modal grain sizes for the cloudy atmospheres are in the \SI{60}{\um} size range (although similar results can be achieved within these models for $\sim$\SI{1}{\um} Fe grains at 1$\%$ super saturation, demonstrating some degeneracies in opacity that depend upon composition and grain size). Due to the red nature of HD~206893~B, we focus on the Cloudy-AE grid, which has the greatest vertical extent in cloud depth; this generally serves to suppress emergent flux at shorter wavelengths and produce higher fluxes at longer wavelengths for the same effective temperature. However, we find that the best fit model incorporating covariances corresponds to low surface gravity (log($g$)=3.5) and the hottest temperature (1700~K) of the four grids, but is remarkably consistent across the four separate band-to-band fits. The full spectral fit corresponds to a model which is significantly bluer than that of HD~206893~B, with the largest estimated corresponding radius (1.55~R$_\textnormal{Jup}$).

\textbf{\textit{BT-Settl Models}:}
The wide range of potential masses and temperatures for the companion estimated by the other model grids motivates further spectral comparison with the BT-Settl (2015) models \citep{allard12, allard14}, which cover larger ranges in these values. BT-Settl was designed to model the atmospheres of very low-mass stars, brown dwarfs, and planets using the PHOENIX radiative transfer atmosphere models \citep{hauschildt92} in conjunction with updated molecular line lists for water, methane, ammonia, and \ce{CO2}, and revised solar abundances derived from the solar radiative hydrodynamical simulations from \citet{asplund09}. The BT-Settl models include a cloud treatment involving supersaturation, which accounts for cloud growth and mixing, and aims to reproduce the stellar M-L and brown dwarf L-T transitions. To conduct a comparison with the HD~206893~B spectra, and given the likely solar metallicity of the host star, we use the solar metallicity BT-Settl (2015) models with grid parameters ranging from 1200~K to 2050~K and surface gravities ranging from log($g$)=2.5--5.5. The resulting best-fit full spectral model shown in Figure~\ref{fig:all_best_models} corresponds to T$_\textnormal{eff}$ of 1600~K, a low surface gravity (log($g$)=3.5), and a radius of 0.96~R$_\textnormal{Jup}$. Along with the best-fit DRIFT-PHOENIX model, this fit has the lowest reduced $\chi^{2}$ incorporating spectral covariance. This model reproduces flux over the $YJ$ bands well, and most closely approximates the $L'$ photometry of all four model suites, but is insufficiently red and departs significantly from the $K$ spectra.

\bibliography{refs_hd206893.bib}


\bibliographystyle{aasjournal}



\end{document}